\documentclass[11pt]{article}  
\usepackage{rotating}
\usepackage{epsfig}
\begin{document}
\title{Strongly-correlated crystal-field approach to 3d oxides \\
- the orbital magnetism in 3d-ion compounds}
\author{R. J. Radwanski and Z. Ropka\\{~}\\Institute of Physics, Pedagogical University, 30-084 Krakow,
POLAND\\
\& Center for Solid State Physics, S$^{nt}$ Filip 5, 31-150 Krakow, POLAND\\{~}\\sfradwan@cyf-kr.edu.pl\\
http:$\backslash$$\backslash$www.css-physics.edu.pl}
 \maketitle
\pagestyle{myheadings} \thispagestyle{plain} \markboth{R.J.
Radwanski and Z. Ropka}{R.J. Radwanski - Strong Crystal Field
Approach} \setcounter{page}{1} \noindent \textwidth = 13.97 cm
dedicated to John H. Van Vleck and to the 75$^{th}$ year
anniversary of the crystal-field theory

\begin{abstract}
We have developed the crystal-field approach with strong electron
correlations, extended to the Quantum Atomistic Solid-State theory
(QUASST), as a physically relevant theoretical model for the
description of electronic and magnetic properties of 3d-atom
compounds. Its applicability has been illustrated for LaCoO$_{3}$,
FeBr$_{2}$ and Na$_{2}$V$_{3}$O$_{7}$. According to the QUASST
theory in compounds containing open 3$d$-/4$f$-/5$f$-shell atoms
the discrete atomic-like low-energy electronic structure survives
also when the 3d atom becomes the full part of a solid matter.
This low-energy atomic-like electronic structure, being determined
by local crystal-field interactions and the intra-atomic
spin-orbit coupling, predominantly determines electronic and
magnetic properties of the whole compound.

We understand our theoretical research as a continuation of the
Van Vleck's studies on the localized magnetism. We point out,
however, the importance of the orbital magnetism and the
intra-atomic spin-orbit coupling for the physically adequate
description of real 3d-ion compounds and 3d magnetism. Our studies
clearly indicate that it is the highest time to ''unquench'' the
orbital moment in solid-state physics in description of 3d-atom
containing compounds.

{~}
{~}
\end{abstract}
\newpage
\maketitle
\tableofcontents
\newpage
\begin{center}
\LARGE{Strongly-correlated crystal-field approach to 3d oxides \\
- the orbital magnetism in 3d-ion compounds\\} \normalsize {~}\\
R. J. Radwanski and Z. Ropka\\{~}\\Institute of Physics,
Pedagogical University, 30-084 Krakow,
POLAND\\
\& Center for Solid State Physics, S$^{nt}$ Filip 5, 31-150 Krakow, POLAND\\{~}\\sfradwan@cyf-kr.edu.pl\\
http:$\backslash$$\backslash$www.css-physics.edu.pl
\end{center}
\section{Introduction}

An unexpected discovery of high-$T_{c}$
superconductivity in 3d-ion oxides in 1986 has revealed the
shortcomings of our understanding of the 3d magnetism. Today we do
not understand the superconductivity but still we do not have a
consistent understanding of electronic and magnetic properties of
3d-ion containing compounds. Despite of very different theoretical
concepts there is no consensus how to treat electrons in unfilled
shells. Many of 3d-ion oxides belong to the class of compounds
called the Mott insulators - they exhibit the insulating state
despite of the unfilled $d$ shell. The fundamental controversy
''how to treat the d electrons'' starts already at the beginning -
should they be treated as localized or itinerant. Directly related
to this problem is the structure of the available states: do they
form the continuous energy spectrum like it is in the band
picture, schematically shown in Fig. 1, or do they form the
discrete energy spectrum typical for the localized states. The
standard band picture encounters serious difficulties - it often
predicts the metallic state for systems that are in the reality
very good insulators, for instance for La$_{2}$CuO$_{4}$, CoO and
NiO \cite{1,2,3,4}. Characteristic feature of the present
literature description of the 3d-ion magnetism is the spin-only
description with the neglect of the orbital magnetism. This
erroneous, according to us, view is related to the widely-spread
conviction about the quenching of the orbital moment in 3d-atom
compounds. This observation made in 1932 by Van Vleck is valid,
however, only in the first-order approximation and for that time
it was a very important theoretical result - it explained
high-temperature dependence of the paramagnetic susceptibility
with the spin-only effective moment. From year of 1996 we advocate
that it is the highest time in the 3d solid-state physics to
''unquench'' the orbital moment and to take into account the
spin-orbit coupling \cite {5} because the spin-orbit coupling is
well-known physical effect and its neglect causes unphysical
oversimplification of the low-energy electronic structure. With
very big surprise we found that the present magnetic community is
not much enthusiastic about it, but we are fully convinced that
the modern description of the magnetism and the electronic
structure have to be made in the atomic scale. One of the reason
for this lack of enthusiasm, assuming that it is only a scientific
sceptism, can be a fact that we unquench the orbital moment with
the use of the crystal-field concept that is disdainfully treated
by present great solid-state theoreticians (experimentalists are
widely using the CEF theory and the atomic-like picture being
pragmatic), though there is rapidly growing experimental evidence
for its ability to describe exceptionally well many physical
properties as it was already shown in numerous papers of Van Vleck
beginning from 1929. We are fully aware of the wide overwhelmed
critics of the crystal-field approach, started with a paper of
Slater in 1953, but we are convinced that this critics is largely
improper. However, we have to say that there is many mistakes,
misuses and oversimplifications of the crystal field in the
present scientific literature, even in the most prestigious
physical journals. A reason for these misuses we think is related
to lack of the open scientific discussion and, say, a
discrimination of the localized approaches in the
presently-in-fashion solid-state theories. In fact, there is
enormous gap between localized picture discovered everyday by
experimentalists and most theoretical descriptions preferring
itinerancy of 3d electrons.

In order to illustrate better the physical problems in description
of 3d-atom containing compounds let present different electronic
structures discussed in recent theoretical approaches to
LaCoO$_{3}$, for instance. This compound has been studied by more
than 50 years due to its nonmagnetic ground state and unusual
temperature dependence of the paramagnetic susceptibility
$\chi$(T) drastically violating the Curie-Weiss law \cite {6,7}
with a pronounced maximum at about 100 K. Fig. 1 presents result
of band-theory calculations for LaCoO$_{3}$ \cite
{8,9,10,11,12,13} and FeBr$_{2}$ \cite {14}- 3d states form
continuous energy spectrum wide in the energy by more than 15 eV.
Fig. 2 presents oversimplified crystal-field schemes from recent
publications \cite {9,15,16}- we recognize these schemes as
single-electron states with weak correlations. In contrary to
band-theory results figure 2 suggests the existence of discrete 3d
states but the electronic structure is only very schematic.

\begin{figure}[!ht]
\begin{center}
\includegraphics [width=11 cm , bb= 20 20 592 356]{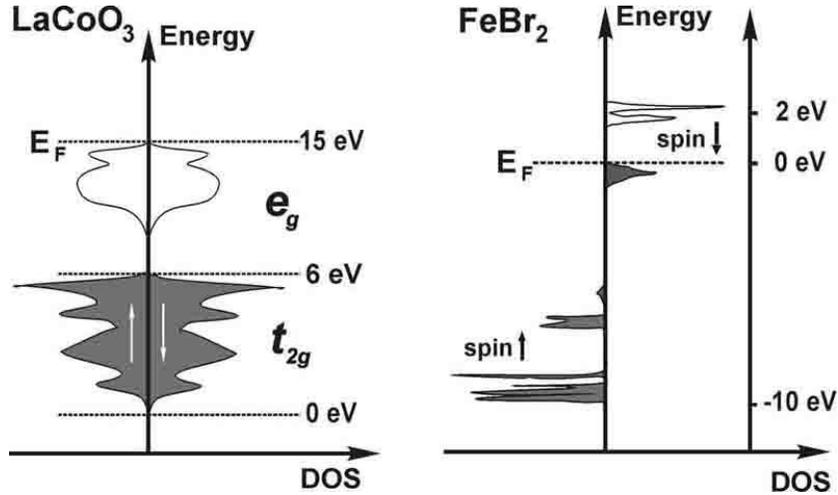}
\caption{Schematic description of the d states in LaCoO$_{3}$ within the band approach \cite{8,9,10,11,12,13} obtained
within the LSDA approach - there is the continuous energy spectrum. There is also some contribution from 2$p$ states of
oxygen. A slightly different picture is obtained within the Hartree-Fock approximation, but still $d$ states form the
continuous energy spectrum spread over 15 eV. Right: spin polarized d states for FeBr$_{2}$ \cite {14}. In both
calculations there is the continuous energy spectrum.}
\end{center}
\end{figure}

\begin{figure}[!ht]
\begin{center}
\includegraphics [width=13.5 cm , bb= 20 20 592 230]{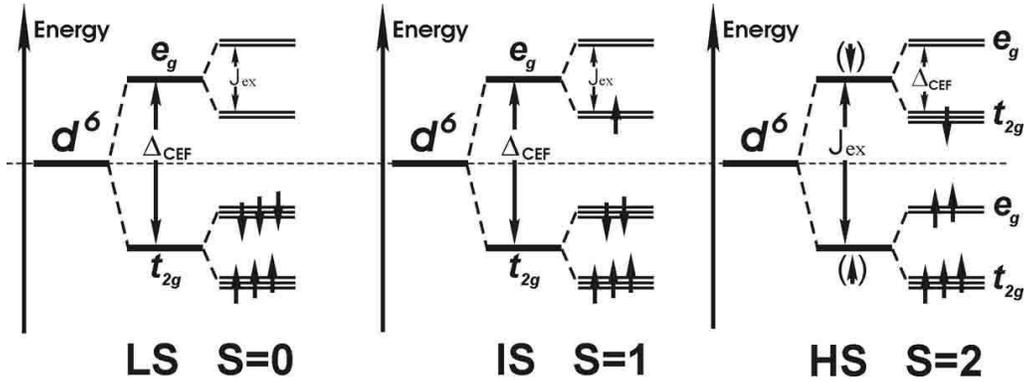}
\caption{Literature schemes of the single-electron discrete energy spectrum of the Co$^{3+}$ ion in LaCoO$_{3}$ for the
low-spin (LS, S=0), intermediate-spin (IS, S=1) and the high-spin state (HS, S=2) realized on the five-fold degenerate
atomic 3d orbitals with lower $t_{2g}$ and higher $e_{g}$ states \cite{8,15}. According to the current literature these
spin states are subsequently realized with the increasing temperature, however, there is no reasonable explanation for
so drastic change of $\Delta_{CEF}$ and $J_{ex}$ parameters with temperature. According to the present approach these
schemes are wrong.}
\end{center}
\end{figure}

\begin{figure}[!ht]
\begin{center}
\includegraphics [width=8.1 cm , bb= 20 20 474 772]{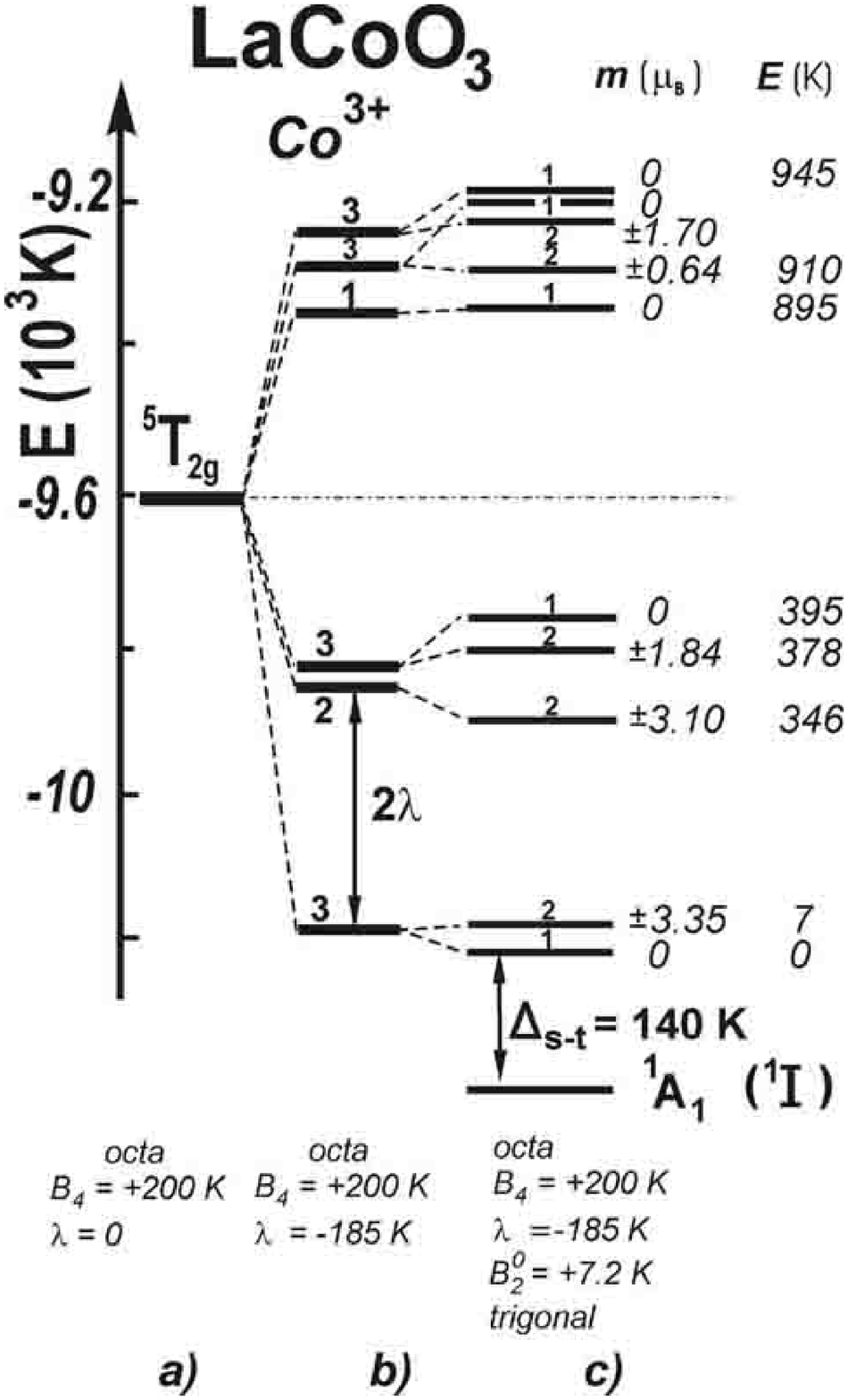}
\caption{Calculated lowest part of the fine electronic structure of the Co$^{3+}$ ion in the CoO$_{6}$ octahedron,
related to the $^{5}$D term, in the presence of the intra-atomic spin-orbit coupling. (a) the cubic $^{5}$T$_{2g}$
subterm is shown only; $^{5}$E$_{g}$ state is at E=+14 400 K, i.e. 2.1 eV above. b) the effect of spin-orbit coupling
on the subterm $^{5}$T$_{2g}$. c) the further splitting by the trigonal distortion. According to the presented
strongly-correlated CEF approach these spin states are increasingly populated with the increasing temperature. The
degeneracy, the magnetic moment and the relative energy of the states are shown. In LaCoO$_{3}$ 140 K below the
structure related to the $^{5}$D term lies, as shown in (c), a non-magnetic singlet $^{1}$A$_{1}$ originating from the
$^{1}$I term. In FeBr$_{2}$ very similar structure exists for the Fe$^{2+}$ ion for the $^{5}$D term like shown in (c),
but without the $^{1}$A$_{1}$ term.}
\end{center}
\end{figure}

The general shape of the bands presented in Fig. 1 can be
understood knowing the localized states of Fig. 2. The continuous
energy spectrum in the band picture looks like a smooth
convolution on the available localized single 3d electron orbitals
$t_{2g}$ (occupied) and higher $e_{g}$ orbitals (empty). The
similarity of the band density of states and the energy level
scheme of Fig. 2 is related to the single-electron treatment of
the 3d electrons in both approaches. In Fig. 3 we present the
crystal-field states with strong electron correlations calculated
by us - here the discrete energy states are many-electron states
of the full Co$^{3+}$ ion \cite{17,18}. In our approach we treat
3d electrons in the incomplete 3d shell as a forming atomic-like
strongly-correlated system. Our strongly-correlated crystal-field
approach for the single 3d cation has been extended to the Quantum
Atomistic Solid State Theory (QUASST) \cite {19,20} in order to
describe 3d-atom compounds.

The revealed fundamental differences in the presented electronic
structures are very important because this low-energy electronic
structure determines physical properties of a real system. We
claim that the states calculated by us are becoming to be more and
more experimentally observed confirming our approach. The
existence of such different electronic structures obviously proves
the lack of a consensus in the understanding of 3d-atom compounds.
There is general agreement that unusual magnetic and electronic
properties of 3d-atom containing compounds and a failure of the
standard band calculations are related to the improper treatment
of electron correlations, that are apparently strong. Thus, the
problem about 3d-atom compounds can be formulated as "how to
account for strong correlations". However, the meaning of strong
correlations is very unclear. In literature it is not clear if an
author means on-site or inter-site electron correlations and which
electrons are correlated. In an one gram of a solid there is
$10^{22}$ atoms and say, 25 times more electrons. We feel that the
strong electron correlations and strongly-correlated electron
systems have not well defined physical meaning and are recalled in
case of non-conventional magnetic and electronic properties. In
fact they are recalled to hide our shortage of the knowledge and
the understanding but the explanation sounds scientifically. With
time we see, attending the annual Strongly-Correlated Electron
Systems (SCES) Conferences and their Proceedings published in
Physica B, that practically all compounds with 3d-/4f-/5f-
open-shell atoms become at present classified as
strongly-correlated electron systems. Surely all 3d oxides are
classified as strongly-correlated electron systems.

Finally, one should note the energy scale on Fig. 3, in particular
that of the first excited state at 7 K (=0.6 meV=5 cm$^{-1}$). It
is at least 1000 times more detailed calculations than the present
band calculations. By present computers we can perform such
detailed calculations, but the problem is whether such states and
such tiny energy separations are preserved in a solid. Developing
the CEF theory and tracing its effects by last 20 years we have
been believing in the substantial applicability of the CEF theory
to solid compounds. We have evaluated a detailed electronic
structure in rare-earth compounds with separations below 0.5 meV
(ErNi$_{5}$ and NdNi$_{5}$ (both are intermetallics) \cite{21},
Nd$_{2}$CuO$_{4}$ (ionic) \cite{22}) but application of the CEF
theory to rare-earth compounds is more acceptable in the magnetic
community. The usefulness of the CEF theory to 3d compounds, even
ionic, has been questioned in presently-in-fashion theories but
such the detailed energy level scheme, as presented in Fig. 3, has
been recently revealed by electron-spin-resonance experiment to
exist in LaCoO$_{3}$ \cite{18,23} confirming the presented
theoretical approach. In doing our research we fully agree with a
Max Planck saying, that "Experiments are the only means of
knowledge at our disposal. The rest is poetry, imagination."

\section{The Aim}

Very general aim of our long-lasting research is to describe
macroscopically-observed physical properties of compounds
containing open shell atoms, in particular with the 3d shell. Some
our research on rare-earth systems one can found in Ref.
\cite{21,24,25} where we successfully applied strongly-correlated
crystal-field approach both for intermetallic and ionic compounds.

In this chapter we would like to present the strongly-correlated
crystal-field approach for the description of electronic and
magnetic properties of 3d-atom compounds, with taking into account
the relativistic spin-orbit coupling. The strong electron
correlations we account for by considering intra-site electron
correlations at a given atom to be energetically dominant. These
strong correlations assure the atomic-like integrity of the
considered 3d atom in the well-defined ionic state even when this
paramagnetic atom becomes the full part of a solid. We named this
approach Quantum Atomistic Solid-State (QUASST) theory as its
starting point for the analysis of a solid containing 3d atoms is
the analysis of constituting atoms and their quantum discrete
electronic structures in a given solid. Here we confine our
discussion to insulating 3d compounds - they are known as Mott
insulators. It was Mott who pointed out that 3d oxides are
insulators despite of having the open shell, violating in this way
the conventional band theory of Wilson and Bloch. The mentioned
above LaCoO$_{3}$ is an example of Mott insulator belonging to the
wide class of oxides with the perovskite structure. As an
exemplary application of QUASST to real compounds our analysis of
FeBr$_{2}$ and ErNi$_{5}$ can be advised \cite {17,18,19,21}. The
former is an ionic compound whereas the latter is an intermetallic
compound.

We do not say that other mechanisms than CEF and the spin-orbit
coupling, after dominant coulombic interactions responsible for
the cohesion of an ionic compound, are not present in a solid, but
we are convinced that analysis of electronic and magnetic
properties of 3d-/4f-/5f-atom compounds is necessary to start from
the determination of the crystal-field interactions and of the
localized CEF-like states.

We understand our theoretical research as a continuation of the
Van Vleck's studies on the localized magnetism. We make use of
efforts of many physicists and chemists of the solid state physics
in order to construct a consistent understanding of electronic and
magnetic properties of 3d-/4f-/5f-atom containing compounds and to
correlate the macroscopically-observed properties with the atomic
scale electronic structure.

\newpage
\section{General idea of understanding of 3d oxides}

\subsection{General idea of QUASST - preservation of the atomic-like discrete electronic structure in a
solid}

A key idea of QUASST is that the paramagnetic ion in a solid
largely preserves its atomic-like integrity, and consequently the
atomic-like electronic structure, when it becomes the full part of
a solid. This atomic-scale low-energy electronic structure
predominantly determines the macroscopically observed magnetic and
electronic properties of the whole compound. As we are interested
in physical properties at room temperature and below we have to
have the electronic structure to be determined below 25 meV (=300
K) with the accuracy better than 1 meV. QUASST is based on
physical concepts well-known in solid-state physics, chemistry and
material science. In QUASST we incorporate the CEF approach, the
group theory, atomic physics, statistical physics, thermodynamics,
the consideration of the local symmetry and many others forming a
consistent theory for understanding of macroscopically-observed
magnetic, electronic and spectroscopic properties of
3d-/4f-/5f-atom compounds in connection to the atomic-scale
low-energy electronic structure of the involved atoms/ions. One
can recognize in QUASST traces of an ionic model, but we employ
the crystal-field approach with strong correlations like shown in
Fig. 3. It contrasts single-electron crystal-field schemes, shown
in Fig. 2, considered in the current literature. It turns out that
most of physics of 3d magnetism lies in details of the low-energy
electronic structure. The atomic-like integrity of the 3d
electrons is pointed out by the writing of 3d$^{n}$
configurations.

\subsection{Formation of a solid}

During the formation of a solid, say an oxide, there proceeds a
charge transfer from the paramagnetic cation to oxygen caused by
the large chemical reactivity of oxygen and its large ability to
capture two electrons from its surroundings. In case of NiO, for
instance, during the formation of the crystal two electrons from
outer 4s+3d shells transfer to the oxygen completing its 2p shell.
Thus we have the charge distribution written as Ni$^{2+}$O$^{2-}$.
An analysis of magnetic and electronic properties of NiO we start
from the discussion of properties of the Ni$^{2+}$ ion in the
crystalline solid as the oxygen has the closed 2p shell
(2p$^{6}$). We do not expect low-energy states related to the
oxygen. Our interest concentrates on the low-energy electronic
structure of the Ni$^{2+}$ ion - in the Ni$^{2+}$ ion there are 8
electrons which can be in the configuration 3d$^{7}$4s$^{1}$,
3d$^{6}$4s$^{2}$, 3d$^{8}$ or a more complex depending on the
invention of a scientist. There is a wide chemical evidence that
with completing the atomic shells with electrons 3d shell is
filled after the 4s shell whereas during the formation of a
compound the 4s electrons are given back the first. So, thinking
about the Ni$^{2+}$ ion in oxides we start with the 3d$^{8}$
configuration. The previous two configurations can lead to more
metallic states, not realized in NiO.

In LaCoO$_{3}$, owing to its insulating ground state, we expect
the charge ion valences as La$^{3+}$Co$^{3+}$O$_{3}$$^{2-}$. In
case of an intermetallic compound like ErNi$_{5}$ or NdNi$_{5}$
the situation is more complex but they can be analyzed as there
localized electrons and itinerant electrons coexist. In case of
ErNi$_{5}$ or NdNi$_{5}$ it has been proved that magnetic
properties are related to the 4f$^{11}$ and 4f$^{3}$
configurations like in the Er$^{3+}$ and Nd$^{3+}$ ions,
respectively. Very good reproduction of the zero-temperature
moment, overall temperature dependence of paramagnetic
susceptibility $\chi$(T) and of heat capacity c(T), including the
magnetic transition proves the existence of the discrete
atomic-like electronic structure with separations below 1 meV
\cite {21,25}.

\subsection{Atomic physics - the existence of atomic terms as
an effect of strong electron correlations}

The 3d$^{8}$ configuration, according to the atomic physics, has
${10 \choose 2} = 45$ states grouped in the Russell-Saunders terms
(the realization of the so-called LS coupling): term $^{3}$F with
21-fold degeneracy, $^{1}$D (5), $^{3}$P (9), $^{1}$G (9) and
$^{1}$S (1). The Hund's rules ground term is $^{3}$F term, that is
described by quantum numbers of the whole 3d$^{8}$ configuration
S=1 and L=3. These terms in a crystal are split and shifted,
sometimes with the energetical reorganization, by the presence of
the crystal-field potential but the integrity of the atomic-like
3d$^{8}$ system is assumed in QUASST to be preserved, at least at
the start. In our description of the ion in a solid we start from
the knowledge collected in the atomic physics, in particular in
respect to the energy of terms and the intra-atomic spin-orbit
coupling. Here the data of NIST Atomic Spectra Database are of the
great importance \cite {26}.

For the 3d$^{6}$ configuration 210 states are grouped in the
atomic terms: $^{5}$D (25 states), $^{3}$H (33), $^{3}$G (27),
2$\cdot$$^{3}$F ($2\cdot21$), $^{3}$D (15), 2$\cdot$$^{3}$P
(2$\cdot$9), $^{1}$I (13), 2$\cdot$$^{1}$G (2$\cdot$9) $^{1}$F
(7), 2$\cdot$$^{1}$D (2$\cdot$5) and 2$\cdot$$^{1}$S (2$\cdot$1).
The Hund's rules term is 25-fold degenerate $^{5}$D term, that is
described by quantum numbers of the whole 3d$^{6}$ configuration
S=2 and L=2.  It is realized as the ground term in case of the
free Fe$^{2+}$ and Co$^{3+}$ ions.

\subsection{Effect of the octahedral crystalline electric field on atomic
states}

In a solid the states of the free Ni$^{2+}$ ion are split and
shifted. In a crystalline solid there is the electrostatic
potential due to other charges forming the solid. This
electrostatic potential is known as the crystal electric field and
its effect on the paramagnetic ion is similar to the Stark effect
in atomic physics. However, the crystal field in solids has very
multipolar character (quadrupolar, octupolar, ...) due to the
regular three dimensional arrangement of charges in a solid. This
multipolar character of the CEF potential is hidden in the CEF
parameters B$_{n}^{m}$. The predominant CEF potential in 3d oxides
is the octupolar term - it is related to the octahedral
crystal-field parameter denoted as B$_{4}$.

In our approach we make use of achievements about the crystal
field reached in analysis of the electron paramagnetic resonance
(EPR) (at present, the name electron spin resonance (ESR) becomes
more popular) on paramagnetic ions put to solids as impurities
\cite {27,28,29,30}. The splitting and the shift of the terms of
3d paramagnetic ions under the action of the cubic crystal field
interactions have been calculated by Tanabe and Sugano already 50
years ago \cite {31}. Unfortunately this knowledge is not used in
presently-in-fashion band-like theories of 3d oxides. The
presently-in-fashion theories of 3d oxides generally employ the
strong crystal field approach that breaks completely intra-atomic
correlations and starts the description of n electrons in the 3d
shell as being largely independent. In the one-electron picture
electrons are put subsequently on the five octahedral CEF
orbitals, with 10 places, on the t$_{2g}$ and e$_g$
single-electron states. This subsequent placing is shown in Fig. 4
drawn with the use of Ref \cite{1} Fig.5.11 on p.150; Ref.
\cite{2} p.1049; Refs \cite{15,16,32,33,34,35}.

In QUASST we apply the weak and intermediate crystal field regime,
that does not break intra-atomic correlations, not only to 3d
impurities, like is considered for the ESR analysis, but to 3d
atoms/ions when they become the full part of a solid.

\begin{figure}[!ht]
\begin{center}
\includegraphics [width=9 cm , bb= 20 20 445 772]{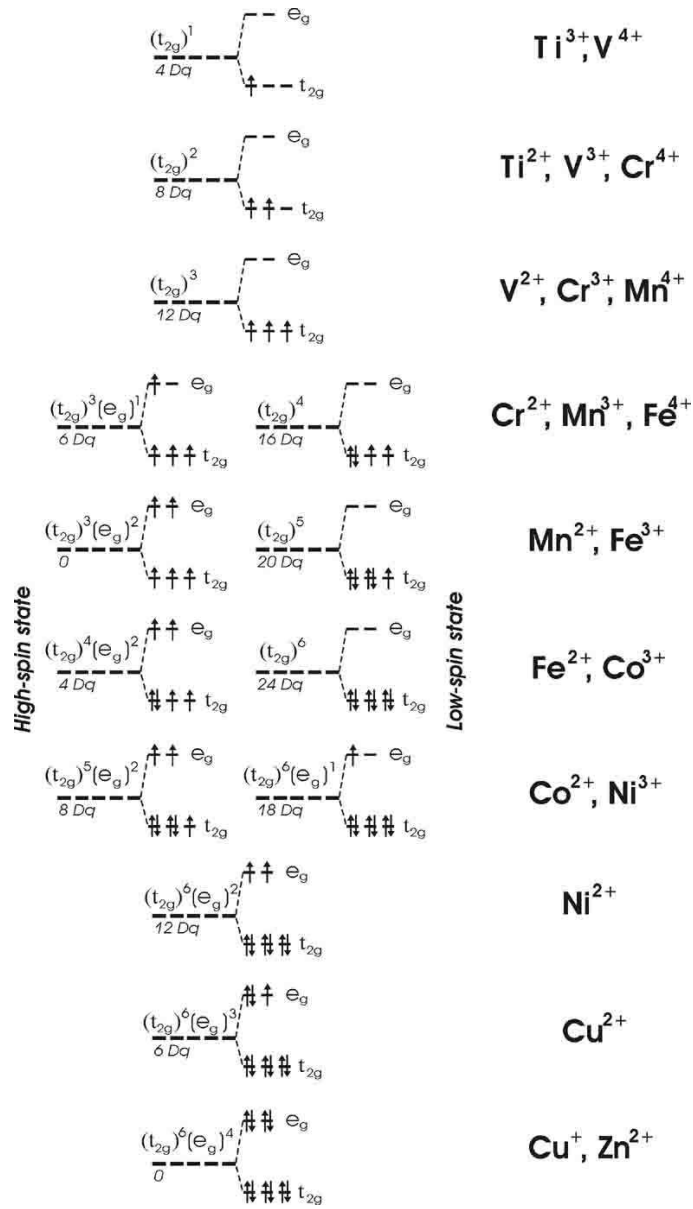}
\caption{Electronic structures of 3d-ions in high- and low-spin states in the octahedral crystal field. Such the
structures are discussed in the present literature \cite{1,2,15,16,32,33,34,35}, but - according to our studies - they
are not physically adequate. Each piece denotes one orbital that can be occupied by two electrons with, up and down,
spin shown.}
\end{center}
\end{figure}

\subsection{Why the octahedral crystal field is the starting
point for discussion of 3d oxides?}

Analysis of the crystallographic structure of many oxides reveals
that the 3d cation is often surrounded by 6 oxygens forming almost
regular octahedron or rarer by 4 oxygens forming a tetrahedron.
The NaCl structure of monoxides NiO or FeO is built up from the
face sharing oxygen octahedra. The perovskite structure of
LaMnO$_{3}$ or LaCoO$_{3}$ is built up from the corner sharing
octahedra. The oxygen octahedra, with the Cu cation sitting
inside, occur also in La$_{2}$CuO$_{4}$, a parent compound of
high-temperature superconductivity materials. Thus the customary
discussion of crystal-field effects in 3d compounds as related to
the octahedral crystal field is largely justified but already here
we would like to say that for the fully physically adequate
description of 3d-atom compounds lattice distortions are very
important and {\bf all}, required by the local symmetry, terms in
the crystal-field potential have to be taken into account.

The effect of the octupolar crystal-field interactions on the
energy states of the Hund's rules ground term is shown in Figs 5b
and 6b. The calculated states are states of the 3d$^{n}$ electron
configuration in the octahedral crystal field produced by the
oxygen octahedron. In case of the tetrahedron the effect of the
octupolar crystal field leads to the reversed schemes.
\begin{figure}[!ht]
\begin{center}
\includegraphics [width=12 cm , bb= 20 20 592 647]{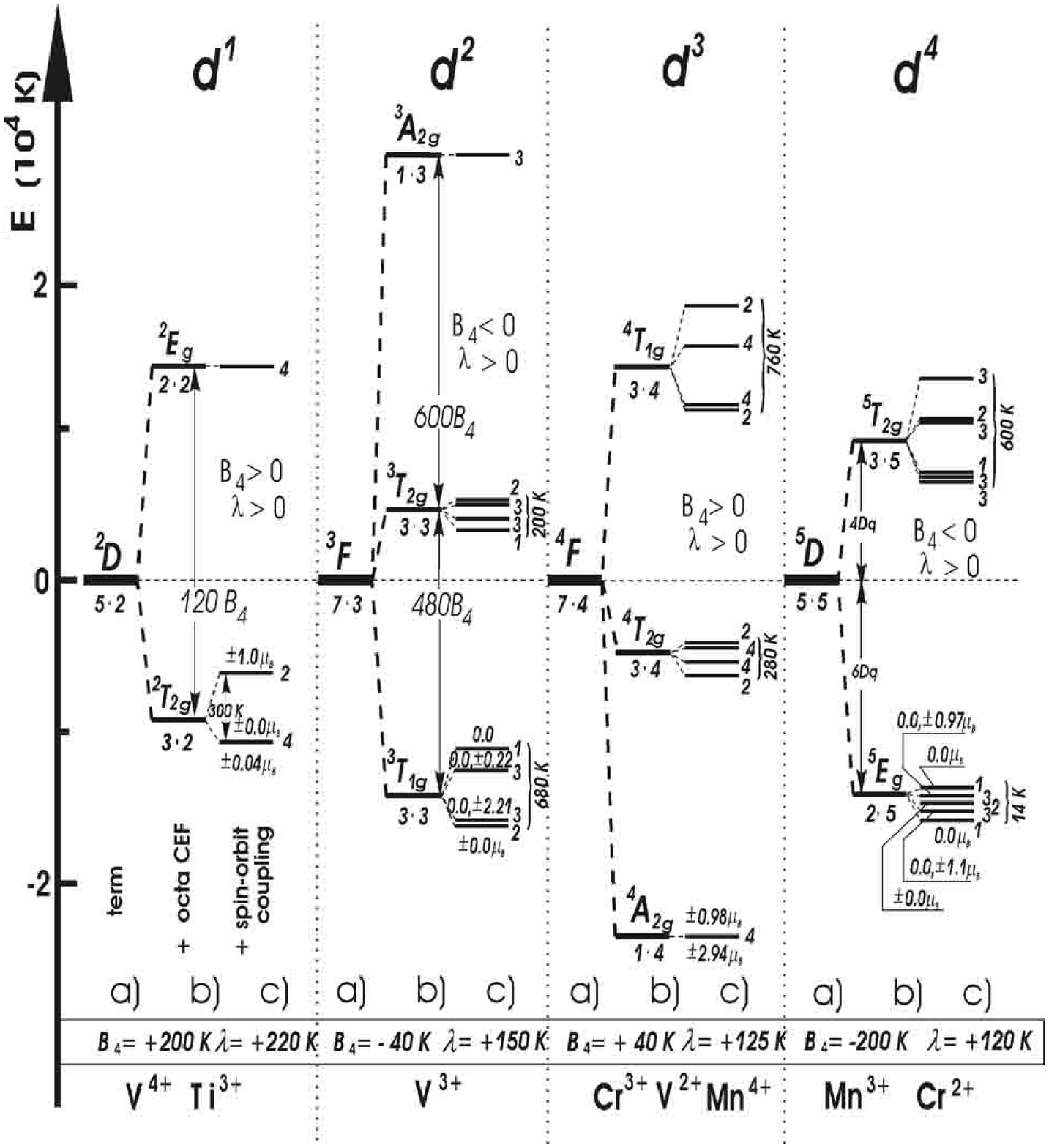}
\caption{The calculated electronic structure of the 3d$^{n}$ configurations of the 3d ions, 1$\leq n$ $\leq 4$, in the
octahedral crystal field (b) and in the presence of the spin-orbit coupling (c). According to the Quantum Atomistic
Solid-State theory the atomic-like electronic structures, shown in (c), are preserved also in a solid. (a) - shows the
Hund's rule ground term. Levels in (c) are labeled with degeneracies in the LS space whereas in (b) the degeneracy is
shown by the orbital spin degeneracy multiplication. The spin-orbit splittings are drawn not in the energy scale that
is relevant to CEF levels shown in figures b. On the lowest localized level the magnetic moment (in $\mu _{B}$~) is
written. The shown states are many electron states of the whole d$^n$ configuration. At zero temperature only the
lowest state is occupied. The higher states become populated with the increasing temperature.}
\end{center}
\end{figure}

\subsection{Can CEF parameters be calculated \emph{ab initio}?}

Having charge distribution in the elementary cell and consequently
in the whole crystal, CEF parameters can be calculated by formulas
known from electrostatics as the respective multipolar moments of
the surrounding charge distribution. By symmetry of the local
surrounding only few multipolar moments are not zero. For
instance, in case of the ligand (=negative ions) octahedron the
only multipolar moment at the center of the octahedron is the
octupolar lattice moment, denoted as CEF coefficient A$_{4}^{0}$,
which in cooperation with the octupolar ion moment provides the
CEF parameter B$_{4}^{0}$.

\subsection{Spin-orbit coupling}

The relativistic spin-orbit coupling is usually ignored in
calculation of the electronic structure of the 3d ions in solids
basing on a general consensus that the s-o coupling is for 3d ions
relatively weak. We have, however, argued \cite{5} that such a
thinking is incorrect - even weak s-o coupling causes dramatic
change of the electronic structure by producing the fine
electronic structure with a large number of low-energy states with
separations even so small as 1 meV (=11.6 K = 8.0 cm$^{-1}$). One
can be surprised, but the influence of the spin-orbit coupling has
not been systematically
studied despite of a quite simple form of the s-o Hamiltonian $H_{s-o}$ = $%
\lambda L\cdot S$. We have proved that the spin-orbit coupling has
to be taken into account for any meaningful analysis of electronic
and magnetic properties of 3d-ion compounds \cite{17,18}. The
weaker spin-orbit coupling the more dramatic influence on the
physical properties at low temperatures is produced.

\begin{figure}[!ht]
\begin{center}
\includegraphics [width=10.5 cm , bb= 20 20 592 654]{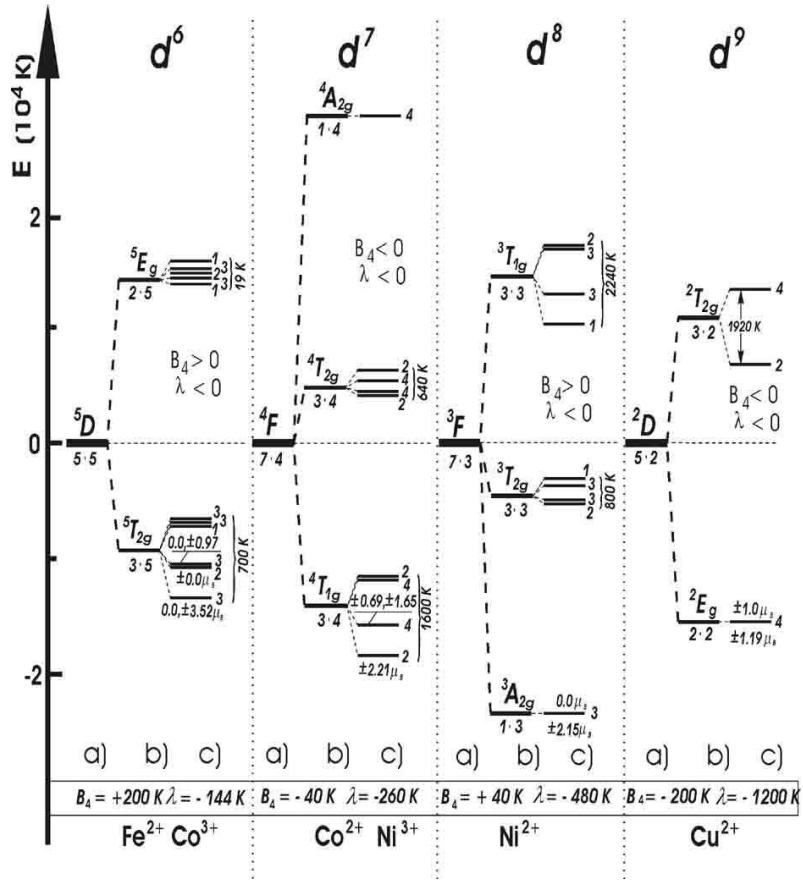}
\caption{The calculated electronic structure of the 3d$^{n}$ configurations of the 3d ions, 6$\leq n$ $\leq 9$, in the
octahedral crystal field in the presence of the spin-orbit coupling. According to QUASST these atomic-like electronic
structures are preserved also in a solid.}
\end{center}
\end{figure}
The influence of the spin-orbit coupling on the localized states
of the strongly-correlated 3d$^n$ electron system produced by
crystal-field interactions of the octahedral symmetry is shown in
Figs 5c and 6c \cite {5,36}.

\subsection{Zero temperature properties, the magnetic state and
thermodynamics}

At zero temperature only the ground state is populated. The
magnetic properties of the compound at T = 0 K are related to this
atomic-like ground state. The states shown in Figs 5c and 6c can
be called the crystal-field or in general charged-formed (CF)
ground state. In dependence of the magnetic characteristics of the
ground state the ion moments can easier or more difficult enter
into the game to create a collective magnetically-ordered state. A
singlet CF ground state, in particular when there is a large
energy separation to the excited states, largely prevents the
formation of the magnetic state. It is worth noting that though
magnetic state is the collective state its formation can be traced
at the single-ion atomic scale as then there appears the breaking
of the time-reversal symmetry in the atomic scale. This symmetry
breaking the easiest is to be observed in case of the CF Kramers
doublet ground state for electronic configurations with an odd
number of electrons (d$^{1}$, d$^{3}$, d$^{5}$, d$^{7}$, d$^{9}$
configurations). The Kramers doublet is only split in the magnetic
state and in presence of the magnetic field, internal or external.
The formation of the magnetic state and a persistent non-magnetic
state will be illustrated by two examples FeBr$_{2}$ and
LaCoO$_{3}$.

With increasing temperature excited states become thermally
populated. The population of states is described by the Boltzmann
statistics. Having the states like those shown in Figs 5c and 6c
the Helmholz free energy F(T) can be calculated by means of
statistical physics. It enables calculations of all
thermodynamical properties like temperature dependence of the heat
capacity c(T), of the paramagnetic susceptibility $\chi$(T), of
the charge quadrupolar ionic moment (measured by the Mossbauer
spectroscopy) and many others. Also temperature dependence of the
magnetic moment in the magnetically-ordered state can be
calculated \cite{17,21}.

QUASST enables calculations of the anisotropy of physical
properties. The paramagnetic susceptibility $\chi$(T), for
instance, can be calculated along different crystallographic
directions.

\section{Hamiltonian, computations and results}
\subsection{Hamiltonian}

In general, the Hamiltonian for the whole compound like NiO or
FeBr$_{2}$ could be written as:
\begin{equation}
H=\sum H_{d} + \sum H_{d-d}
\end{equation}
where H$_{d}$ should treat intra-atomic whereas H$_{d-d}$
inter-atomic interactions. H$_{d-d}$ comprises, for instance, the
energy between Ni$^{2+}$ and O$^{2-}$ ions (called the lattice
Madelung energy), that in case of an ionic compound like NiO is
dominant, but also much, much smaller energy between magnetic
ionic moments leading to the magnetically-ordered state. The
separation for intra and inter-atomic interactions is not strict
as, for instance, the crystal-field energy, that is considered as
the single-ion effect in fact is the coulombic charge multipolar
interactions of the 3d cation with the charge multipolar moments
of the surroundings.

In consideration of H$_{d}$ we assume the physical situation for
the 3d$^n$ system of a 3d-transition-metal ion to be accounted for
by considering the single-ion-like Hamiltonian containing the
electron-electron d-d interactions $H_{el-el}$, the crystal-field
$H_{CF}$, spin-orbit $H_{s-o}$, and Zeeman $H_Z$ interactions:

\begin{equation}
H_d=H_{el-el}+H_{CF}+H_{s-o}+H_Z
\end{equation}

The electron-electron and spin-orbit interactions are intra-atomic
interactions, whereas crystal-field and Zeeman-like interactions
account for interactions of the unfilled 3d shell with the charge
and spin surroundings. These interactions are written in the
decreasing strength succession. Important is that we assume that
H$_{CF} \ll$ H$_{el-el}$.

It is the essentially important to assume the dominant
interactions to be H$_{el-el}$. In fact, here is the clue of our
strongly-correlated crystal-field approach based on the atomistic
approach. H$_{el-el}$ is responsible for the formation of the
atomic terms from ${10 \choose n}$ states. A number of states,
allowed by the Pauli exclusion principle, is large and amounts 10
for d$^1$/d$^9$, 45 for d$^2$/d$^8$, 120 for d$^3$/d$^7$, 210 for
d$^4$/d$^6$ and 252 for the 3d$^{5}$ configuration. The given term
gather states with the same quantum numbers L and S and they are
rather well separated from other terms. The ground term of the
free atom/ion is determined by two phenomenological Hund's rules,
i.e. the lowest term of the whole 3d$^{n} $ system is
characterized by 1$^{o})$ the maximal value of the resultant spin
quantum number $S$ and 2$^{o}$) the maximal value of the resultant
orbital quantum number $L$ provided the condition 1$^{o}$. These
rules yield for, for instance, the 3d$^{4}$ and 3d$^{6}$ electron
configurations, the term $^{5}D$ with $S$ = 2 and $L$ = 2 as the
ground term. This term is 25-fold degenerated in the spin-orbital
space. In a solid other then Hund's rules ground term can be
realized as the ground term, like it occurs in LaCoO$_{3}$ and is
shown in Fig. 3c. Figs 5 and 6 have been calculated for Hund's
rules ground terms.

In QUASST we take into account on-site electron-electron
correlations among d electrons by assuming that also for the ion
in a solid the term structure is preserved. It does not mean at
all that we claim that in a solid the term structure is exactly
the same as in the free ion. The term structure in a solid differs
from that of the free ion but we say that in calculations of the
influence of the crystal-field potential and of magnetic
interactions we should follow along well known from atomic physics
interactions keeping the atomic-like integrity of the 3d$^{n}$
system. The strong intra-atomic correlations allow to work with
many-electron quantum numbers $S$ and $L$ of the whole 3d$^{n}$
configuration instead of single-electron states with $s_{i}$ and
$l_{i}$. In band structure calculations and in the single-electron
crystal-field approach it is assumed that the atomic-like
d-electron correlations have been broken and then the better
starting point is to consider d electrons to be largely
independent but experiencing separately the octahedral crystal
field (H$_{CF}\gg$ H$_{el-el}$). In the consequence in the
single-electron crystal-field approach electrons are put one by
one on the five d orbitals determined by the octahedral crystal
field (three t$_{2g}$ states below and two e$_{g}$ states higher).
In such approach electron correlations are at the beginning absent
and then they are introduced into calculations.

At beginning let assume that in a solid the ground term is the one
given by phenomenological Hund's rules. The Hund's rules ground
term of the 3d$^n$ electron systems are collected in Table 1. The
LS ground term is (2L+1)$\cdot $(2S+1) degenerated. Strong
intra-atomic Hund's-rule correlations allow to work in the
$|LSL{_z}$S${_z}\rangle$ space, though one should remember that
the complete QUASST calculations should be performed in the
all-terms space with ${10 \choose n}$ states. In the one-term
$|LSL{_z}$S${_z}\rangle$ space the effect of the crystal-field, of
the spin-orbit coupling and of the (internal/external) magnetic
field is accounted for by considering the single-ion Hamiltonian
of the 3d$^n$ system of the 3d-transition-metal ion of the form:
\begin{equation}
H_d ^1=\\
\sum\sum B_{n}^{m}\hat{O}_{n}^{m}(L,L_{z}) + \lambda L\cdot S +
\mu_{B} (L+g_{e}S)\cdot B_{ext}
\end{equation}
where $^1$ denotes the one-term Hamiltonian left after the
dominant action of H$_{el-el}$.
The first term is the CEF Hamiltonian with the Stevens operators $%
\hat{O}_n^m $ that depend on the orbital quantum numbers L, L$_z$.
B$_{n}^{m}$ are CEF parameters. The octahedral crystal field
takes, for the z axis along the cube edge, the form
\begin{equation}
H_{CF} = B_4(\hat{O}_4^0 (L, L_z) + 5\cdot\hat{O}_4^4 (L, L_z))
\end{equation}
B$_4$ is the octahedral CEF parameter that depends on the
octupolar charge moment of the 3d shell and the octupolar moment
of the lattice surroundings. The octupolar charge moment of the 3d
shell is known as the Stevens coefficients $\beta$ (Table 1, after
Ref. \cite{27}, p. 873). The second term in Hamiltonian (3)
accounts for the spin-orbit interactions. The last term accounts
for the influence of the magnetic field, the externally applied in
the present case. g$_e$ value is taken as 2.0023.

\subsection{Computations and results}
The computations of the many-electron states of the 3d$^n$ system
have been performed by consideration of the Hamiltonian (3) in the
$|LSL{_z}$S${_z}\rangle$ base. As a result of the exact
diagonalization, of the maximal matrix of 28 by 28, we obtain the
energies of the (2L+1)$\cdot $(2S+1) states and the eigenvectors
containing information e.g. about the magnetic characteristics.
These magnetic characteristics are computationally revealed under
the action of the external magnetic field B$_{ext}$. The
(2L+1)$\cdot $(2S+1) degeneracy is removed by i) the crystal field
(CEF) interactions and ii) by the intra-atomic spin-orbit
coupling. Despite of the fact that for the 3d ions the spin-orbit
coupling is by two-orders of magnitude weaker than the CEF
interactions we do not apply the perturbation method, as is
usually made in literature, but we treat the CEF and spin-orbit
interactions on the same footing. The calculated
electronic structure of the 3d ion with the 3d$^{n}$ configuration, 1$\leq n$ $%
\leq 9$, are collected in Figs 5 and 6 for the octahedral symmetry
of the crystal field.

\begin{sidewaystable}
\begin{center}
\begin{tabular}{|c|c|c|c|c|c|c|c|c|c|c|c|}
  \hline
  $~~$n$~~$ & $~~$S$~~$ & $~~$L$~~$ & \multicolumn{2}{|c|}{free ion} & \multicolumn{4}{|c|}{octahedral crystal-field} & \multicolumn{3}{|c|}{with the spin-orbit coupling} \\
  \cline{4-12}
   &  &  & ground term & degen. & CEF state & degen. &$~~\beta~~$&$~~$B$_4$ (K)$~~$ &$~~\lambda$ (K)$~~$&$~~\xi$ (K)$~~$&degen.\\
  \hline
  & & & & & & & & & & &\\
  d$^1$ & $\frac{1}{2}$ & 2 & $^2$D & 10 & $^2T_{2g}$ & 6 &$~~$+$\frac{2}{63}~~$& +200 & +220 & +220 & 4\\
    & & & & & & & & & & &\\
  d$^2$ & 1 & 3 & $^3$F & 21 & $^3T_{1g}$ & 9 & -$\frac{2}{315}$ & -40 & +150 & +300 & 2\\
    & & & & & & & & & & &\\
  d$^3$ & $\frac{3}{2}$ & 3 & $^4$F & 28 & $^4A_{2g}$ & 4 & +$\frac{2}{315}$ & +40 & +125 & +375 & 4\\
    & & & & & & & & & & &\\
  d$^4$ & 2 & 2 & $^5$D & 25 & $^5E_g$ & 10 & -$\frac{2}{63}$ & -200 & +120 & +480 & 1\\
    & & & & & & & & & & &\\
  d$^5$ & $\frac{5}{2}$ & 0 & $^6$S & 6 & - & 6 & 0 &  & - & - & 6\\
    & & & & & & & & & & &\\
  d$^6$ & 2 & 2 & $^5$D & 25 & $^5T_{2g}$ & 15 & +$\frac{2}{63}$ & +200 & -140 & +560 & 3\\
    & & & & & & & & & & &\\
  d$^7$ & $\frac{3}{2}$ & 3 & $^4$F & 28 & $^4T_{1g}$ & 12 & -$\frac{2}{315}$ & -40 & -260 & +780 & 2\\
    & & & & & & & & & & &\\
  d$^8$ & 1 & 3 & $^3$F & 21 & $^3A_{2g}$ & 3 & +$\frac{2}{315}$ & +40 & -480 & +960 & 3\\
    & & & & & & & & & & &\\
  d$^9$ & $\frac{1}{2}$ & 2 & $^2$D & 10 & $^2E_g$ & 4 & -$\frac{2}{63}$ & -200 & -1200 & +1200 & 4\\
    & & & & & & & & & & &\\
  d$^{10}$ & 0 & 0 & $^1$S & 1 & - & 1 & - &  & - & - & 1\\
  \hline
\end{tabular}
\caption{Spin S and orbital L quantum numbers of the two Hund's
rules ground term for the strongly-correlated 3d$^n$ electron
systems with the total degeneracy in the spin-orbital space. The
ground octahedral subterm and the degeneracy in the octahedral
cubic crystal field with the respective values of the octupolar
CEF parameter B$_4$, that alternates with the Stevens coefficient
$\beta$. The spin-orbit coupling parameter: many-electron $\lambda
$ and single-electron $\xi$; $\lambda $ = $\pm\xi$/2S. The last
column - the degeneracy resultant from the spin-orbit coupling in
the presence of the octahedral cubic CEF interactions. These
values of B$_4$ and $\lambda $ have been used for calculations of
Figs 5 and 6.}
\end{center}
\end{sidewaystable}
Figs 5 and 6 present the calculated general overview of the
octahedral CEF and spin-orbit effect on the Hund's rule term for
the 3d$^n$ systems. Figures c show the splitting of the ground
term by the octahedral cubic CEF interactions in the presence of
spin-orbit interactions. The parameters used are collected in
Table 1.

The calculations have been performed with the realistic octahedral
crystal field parameter B$_{4}$. The $T_{2g}$-$E_{g}$ splitting,
approximately equal to 120$\cdot$B$_{4}$, amounts to 2.1 eV for
the 3d$^{1}$ system in agreement with the quite frequent
experimental observations of the d-d excitations in the optically
visible energy range, i.e. 1.7-3.5 eV. The spin-orbit coupling
values ($\lambda$ = 220-1200 K) are taken for divalent free ions,
see for instance Ref. \cite{27}, p. 399. The octahedral crystal
field strongly dominates the effect of the spin-orbit coupling.
The parameter B$_{4}$ alternates with the Stevens coefficient
$\beta$, that is a measure of the angular part of the octupolar
charge moment of the 3d cloud for the Hund's rules ground term,
Fig. 7. The shown values of B$_{4}$ keep the same lattice
contribution, A$_{4}$, i.e. the same octupolar charge moment of
the surroundings, because we take always the octahedral oxygen
surroundings into account. In the simplest form for the
point-charge model the CEF parameter B$_{4}$ for the given term is
expressed as:
\begin{equation}
B_{4} =\\\beta \cdot <r_{d}^{4}> \cdot A_{4}
\end{equation}
For the octahedral symmetry the CEF coefficient A$_{4}$ is
expressed as \cite{27}, p. 669:
\begin{equation}
A_{4} = -\frac{7}{16} \frac{Ze^{2}}{d^{5}},
\end{equation}
where Z is the charge of oxygen (-2) and d is the cation-oxygen
distance (192.5 pm in LaCoO$_{3}$). It yields a value of A$_{4}$
of +432 Ka$_{B}$$^{-4}$, a$_{B}$ is the Bohr radius. Taking for
the Co$^{3+}$ ion $\beta$ = +2/63 and $<$r$_{d}$$^{4}$$>$=2.342
a$_{B}$$^{4}$ \cite{37} we get B$_{4}$ = +32 K. This value is
eight times smaller than the recent evaluation of B$_{4}$ of 260 K
\cite{18}, but the most important is that 1) this \emph{ab initio}
calculations give the proper sign of the B$_{4}$ parameter as it
determines the ground state in the oxygen octahedron and 2) the
experimentally derived strength of crystal-field interactions
turns out to be much weaker than it was thought in literature for
justification of the strong crystal-field approach. We are not
going to discuss the discrepancy in B$_{4}$ - surely it is
necessary to take into account all charges. Next neighbors, the
cube of La$^{3+}$ ions adds to the nearest-neighbors oxygen
contribution. This adding, despite of different sign of charges of
the La and oxygen ions, is of great importance as there is a hope
that by taking into account all charges the parameter B$_{4}$ will
increase becoming closer to the experimental value. For Figs 5 and
6 we assume B$_{4}$ = +200 K. In a pragmatic approach in the start
we take B$_{4}$, and other CEF parameters, as parameters to be
evaluated from the analysis of properties of a real compound in
frame of the CEF approach with the given symmetry.
\begin{figure}[!ht]
\begin{center}
\includegraphics [width=9 cm , bb= 20 20 592 701]{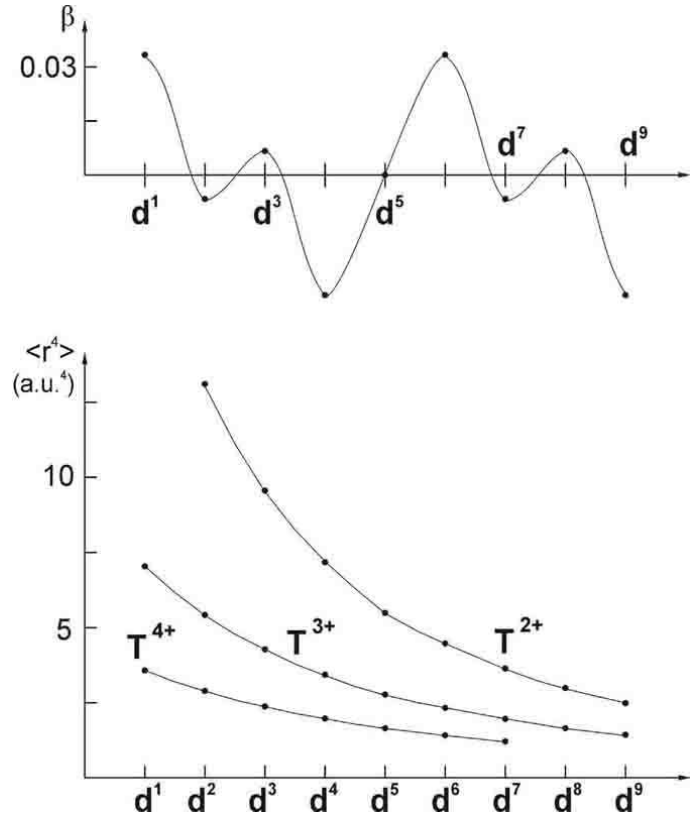}
\caption{Octupolar moment of the 3d$^{n}$ shell in the Hund's rules ground term. a) $\beta$ is the Stevens coefficient
that describes the angular part of the quadrupolar moment. b) $<r^4>$ is the mean values of the radius in power of 4,
in atomic units, for 3d cations in different valence states calculated by Freeman and Watson \cite{37}; here taken from
Ref. \cite {27}, p. 399.}
\end{center}
\end{figure}

Figures 5 and 6 are full of information. One can see the
similarities, but also the differences, between the CEF effect and
s-o interactions on, for instance, the $^2$D and $^5$D terms for
the d$^1$ ($\lambda > 0$) and d$^6$ ($\lambda < 0$)
configurations. The particle-hole symmetry can be studied for the
d$^n$/d$^{10-n}$ configurations. We mention here only a few most
important points.

\begin{enumerate}
\item  For 3d ions there are only D and F terms as L can be only 2
and 3. The spin degeneracy depends on the number of electrons
involved. The total degeneracy of low-energy states as large as 15
is realized.

\item  The pairs d$^1$/d$^6$, d$^2$/d$^7$, d$^3$/d$^8$ and
d$^4$/d$^9$ have the same orbital ground state. The spin-orbit
effect is, however, entirely different due to different values of
S, the reversal of the s-o constant $\lambda $ and the
transformation of non-Kramers ions into Kramers ions (even
\emph{vs} odd number of d electrons).

\item  The adoption (affinity) or the loss of the one electron draws the full
reconstruction of the fine electronic structure.

\item  The spin-orbit coupling removes largely the degeneracy in
all cases apart of the spin-only d$^5$~ system. In case of the
d$^4$ configuration the s-o coupling yields a singlet ground
state, but there are 10 closely lying levels.

\item The s-o coupling produces a fine electronic structure with
allowed excitations even much below 5 meV. These low-energy
excitations can be detected by heat capacity and spectroscopic
methods like Electron Spin Resonance.

\item  There is a tiny splitting of the $^5$E$_g$ cubic subterm realized for the Mn$^{3+}$ ion
in LaMnO$_{3}$, recently a very popular compound.

\item  There is also the extra second-order s-o splitting of the
T$_{2g}$ originated states in comparison to the perturbation
method and to the so-called T-P term equivalence used in Ref.
\cite{29}.

\item  The ground-state magnetic moment substantially differs from the spin-only value, i.e. from the integer
value of 2$n$ $\mu _{B}.$ A very small moment has been revealed
for the d$^1$, d$^2$ and d$^4$ configurations.

\item  The states shown are {\bf many-electron states} of the
whole 3d$^n$ system. At {\bf 0 K only the lowest state is
occupied. Higher excited states become populated with increasing
temperature.}

\item  The population of higher states manifests in e.g. temperature
variation of electronic and magnetic properties like the heat
capacity c(T) and the magnetic susceptibility $\chi$(T). Detailed
calculations of $\chi $(T) will be presented for the d$^6$ system
in FeBr$_{2} $ for the Fe$^{2+}$ ion.
\end{enumerate}

For the physical understanding the most important are the points
9, 8 and 3. They illustrate the fundamental difference with the
very often recalled one-electron picture with subsequent
occupation of the t$_{2g}$ and e$_g$ states, as is shown in Fig.
4, without the reconstruction of the electronic structure in
contrast to the present result of 3. In the many-electron picture
at 0 K only the lowest state is occupied, i.e. the lowest-state
energy is the energy of all n electrons as the whole d$^n$
configuration.

The evaluation of the fine electronic structure, with the given
energy and magnetic characteristics, allows for the calculation of
temperature dependence of many physical properties like heat
capacity or the paramagnetic susceptibility similarly to the
description successfully used for the rare-earth compounds
\cite{19,21,22,25}.

\subsection{Example I: Magnetic Mott insulator: FeBr$_{2}$
(antiferromagnet)}

\subsubsection{Properties and Hamiltonian}

FeBr$_{2}$ exhibits insulating ground state with the charge
distribution Fe$^{2+}$Br$_{2}^{1-}$. It is antiferromagnetic with
T$_{N}$ of 14 K (for references to experimental data see to Ref.
\cite{17}). After the metamagnetic transition at 3 T, it exhibits
magnetic moment of 4.3 $\mu _{B}$ per formula unit, for what we
charge the Fe$^{2+}$ ion. The electronic structure of the
Fe$^{2+}$ ion in the $^{5}$D term is shown in Fig. 8e. It is very
similar to that shown in Fig. 3 for the Co$^{3+}$ ion (Fe$^{2+}$
and Co$^{3+}$ ions are isoelectronic 3d$^{6}$ systems) but the
lowest splitting is opposite, i.e. the doublet is lower than
singlet, see also Fig. 9a.
\begin{figure}[!ht]
\begin{center}
\includegraphics [width=7.7 cm , bb= 20 20 592 744]{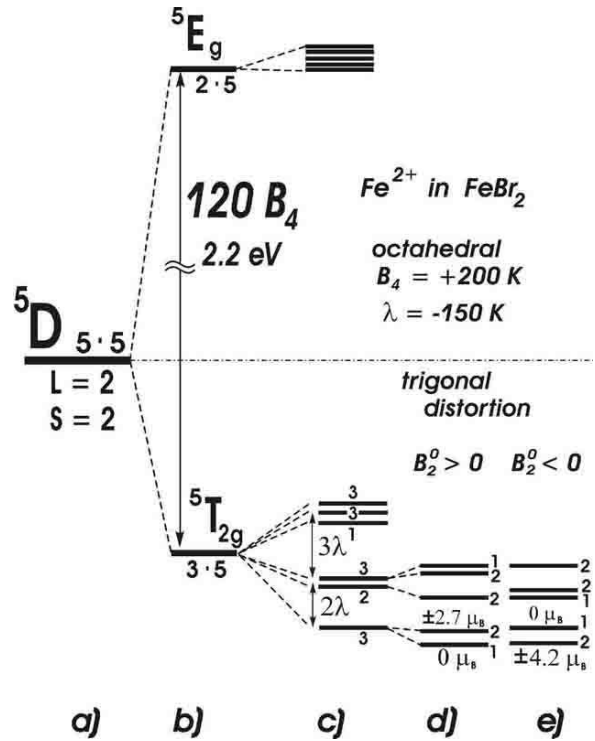}
\caption{Fine electronic structure of the strongly-correlated 3d$^6$ electronic system. a) The 25-fold degenerated
Hund's rules $^5$D ground term. b) the effect of the oxygen octahedral crystal field, c) the effect in combination with
the spin-orbit coupling; d) and e) further splittings due to trigonal distortions - the case e) is realized in
FeBr$_{2}$.}
\end{center}
\end{figure}
\begin{figure}[!ht]
\begin{center}
\includegraphics [width=7.7 cm , bb= 20 20 383 772]{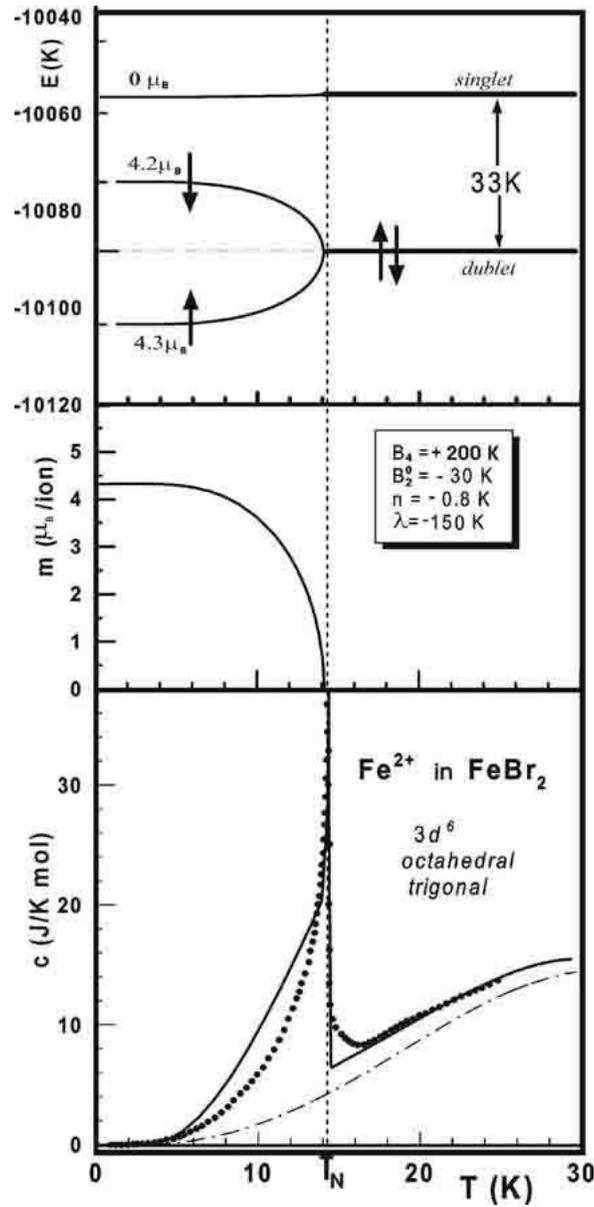}
\caption{\footnotesize{Magnetic phase transition in FeBr$_{2}$. a) The calculated temperature dependence of the lowest
part (3 lowest states) of the electronic structure of the 3d$^{6}$ configuration in the octahedral crystal field
(B$_{4}$=+200 K), in the presence of the spin-orbit coupling ($\lambda$ = -150 K), trigonal distortion B$_{2}^{0}$ =
-30 K and the molecular-field coefficient n = -0.8 K/$\mu _{B}^{2}$. b) temperature dependence of the atomic magnetic
moment of the Fe$^{2+}$ ion; c) Calculated temperature variation of the contribution of the d subsystem to the heat
capacity of FeBr$_{2}$. Points denote literature experimental data.}}
\end{center}
\end{figure}
\begin{figure}[!ht]
\begin{center}
\includegraphics [width=10.7 cm , bb= 20 20 592 501]{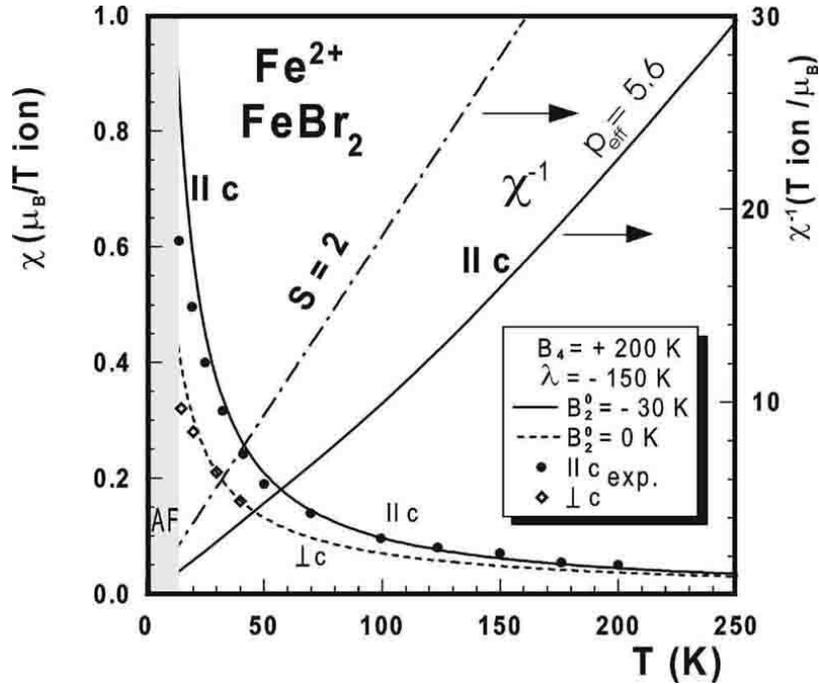}
\caption{Calculated temperature dependence of the paramagnetic susceptibility $\chi$(T) of FeBr$_2$ for magnetic fields
applied along and perpendicular hexagonal c axis. Calculations have been performed for B$_4$=+200 K, $\lambda$$_{s-o}$
= -150 K and the trigonal parameter B$_{2}^{0}$ = -30K. Points are experimental values. A shadow area shows the
antiferromagnetic state below T$_{N}$ of 14.2 K. Curve marked as S=2 shows $\chi$(T) expected from Curie law with S=2
\cite {17}.}
\end{center}
\end{figure}
For the description of the magnetic state we use a Hamiltonian for
the Hund's rules ground term ($S$=2, $L$=2) completed with a
spin-dependent magnetic term \cite{17}:

\begin{equation}
H_{d} ^1=\\
-\frac{2}{3} B_{4} (\hat{O}_{4}^{0}-20\surd{20}\hat{O}_{4}^{3}) +
\lambda L\cdot S+n\left(
-\hat{m}<\hat{m}>+\frac{1}{2}<\hat{m}>^{2}\right) + B_{2}^{0}
\hat{O}_{2}^{0}
\end{equation}
where $\hat{m}$=-($\hat{L}$+2$\cdot$$\hat{S}$) $\mu _{B}$ is the
magnetic-moment operator of the $d$-atom and $n$ is the
molecular-field coefficient. The first term is the octahedral CEF
Hamiltonian written for the z axis taken along the cube diagonal
(then a coefficient -2/3 to B$_{4}$ appears), because FeBr$_{2}$
exhibits trigonal distortion. The third term accounts for
spin-dependent interactions. The last term accounts for the slight
off-octahedral trigonal distortion of the local Br$^{1-}$
octahedron. The magnetically-ordered state is calculated
self-consistently. The optimal parameters found for FeBr$_{2}$
are: the octahedral CEF parameter $B_{4}$ = +200 K, the spin-orbit
coupling $\lambda $ = -161 K, n = -0.8 K/$\mu _{B}^{2}$ and the
trigonal off-cubic distortion $B_{2}^{0}$ = -30 K. The magnetic
ordering temperature occurs for:
\begin{equation}
n = \chi^{-1} (T_{N})
\end{equation}

\subsubsection{Thermodynamics - Counting of atoms}

The calculated temperature dependence of the ordered magnetic
moment and of the heat capacity are shown in Fig. 9 together with
the temperature dependence of the energy of the three lowest
localized states shown in Fig. 8. Everybody admits that the
attained agreement in description of properties of FeBr$_{2}$,
both zero-temperature magnetic moment of 4.3 $\mu _{B}$ and
thermodynamics, is remarkably good. The full reproduction of
temperature dependence of the heat capacity, of the paramagnetic
susceptibility and of the macroscopic magnetization proves that
{\bf all} Fe atoms equally contribute to these properties. It
means that {\bf all} atoms are in the Fe$^{2+}$ ion state with the
same electronic structure. In fact, by the reproduction of values
of the macroscopically-observed properties actually we count
magnetic atoms. For comparison of the calculated microscopic
atomic-scale values and macroscopic molar values simply the
Avogadro number is used only.

So good description is thanks the crystallographic structure (and
the good theory, of course). Thanks the hexagonal structure all Br
octahedra are aligned with the main diagonal along the hexagonal c
axis. Then the trigonal distortion of the local octahedron can
proceed coherently on all Fe sites without breaking of the
hexagonal symmetry of the elementary cell.

\subsubsection{Orbital and spin moments}

Very important superiority of QUASST among the present solid-state
theories lies in the possibility of the calculations of the
orbital magnetic moment. The orbital moment comes out from the
intra-atomic spin-orbit coupling. In FeBr$_{2}$ we have found that
the Fe$^{2+}$ ion moment is composed from the orbital moment
m$_{O}$=+0.80 $\mu _{B}$ and the spin-moment m$_{S}$=+3.52 $\mu
_{B}$ \cite{17}. In NiO with the 3d$^{8}$ configuration of the
Ni$^{2+}$ ion we have calculated
the orbital moment m$_{O}$ of +0.54 $\mu _{B}$ and the spin moment m$%
_{S}$ of +1.99 $\mu _{B}$ \cite{38}. The existence of a quite
large orbital moment in NiO and CoO has been revealed recently by
X-ray synchrotron radiation \cite{39,40}- we take this fact as
confirmation of the QUASST theory. Consistent description, keeping
the same lattice octupolar moment, of properties of NiO and CoO
has been presented in Ref. \cite{41}. The same lattice octupolar
moment is desirable owing to the same crystallographic structures.
The attained good description indicates on the dominant single-ion
mechanism.

\subsection{Example II: Nonmagnetic Mott insulator: LaCoO$_{3}$}

LaCoO$_{3}$ attracts a large scientific interest by more than 50
years due to its intriguing non-magnetic ground state and
anomalous temperature dependence of the paramagnetic
susceptibility $\chi$(T) with a pronounced maximum at about 100 K.
This nonmagnetic state is at the atomic scale. It is widely
believed that with increasing temperature the Co$^{3+}$ ion
changes its state from the low-spin (LS) to high-spin (HS) state
via an intermediate (IS) state \cite{9}, as is schematically shown
in Fig. 2. Our studies provide much more exact description of
states, see Fig. 3 \cite{18}, and exclude the existence of the IS
state. Within a HS configuration, described in QUASST by the
$^{5}D$ term with S=2, the Co$^{3+}$ ion can have non-magnetic
singlet ground state as an effect of a lattice off-octahedral
distortion in the presence of the spin-orbit coupling, but in
LaCoO$_{3}$ the singlet ground state is a non-magnetic $^{1}A_{1}$
subterm that originates from the $^{1}I$ term.
\begin{figure}[!ht]
\begin{center}
\includegraphics [width=12 cm , bb= 20 20 592 404]{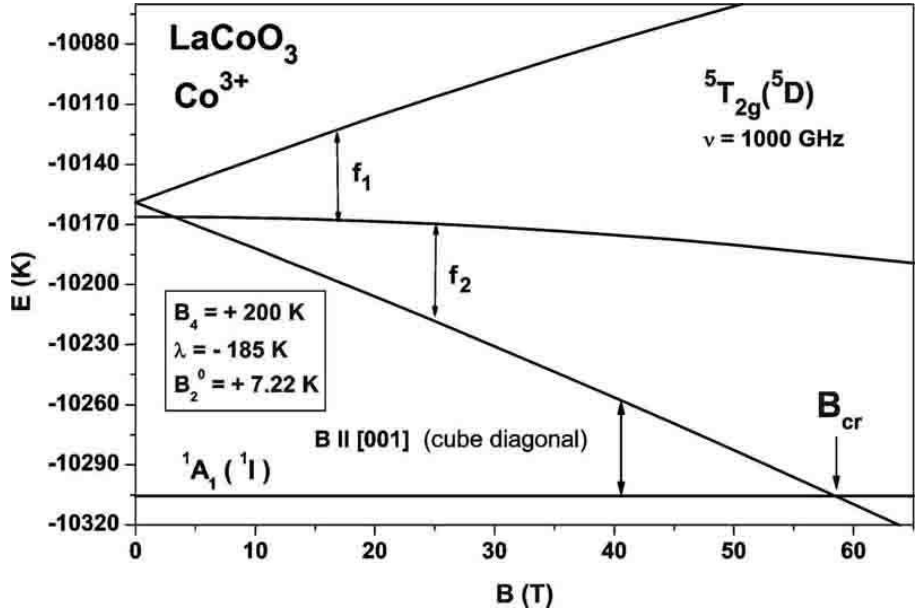}
\caption{Calculated field dependence of the lowest part of the electronic structure of the Co$^{3+}$ ion in LaCoO$_{3}$
for external magnetic fields applied along the diagonal of the cube of the perovskite structure of LaCoO$_{3}$ of the
quasi triplet originating from the 25-fold degenerate $^{5}D$ term for the octahedral crystal field B$_{4}$ = +200 K
and the spin-orbit coupling $\lambda $ = -185 K and the distortion trigonal parameter B$_{2}^{0}$ = +7.2 K. The ESR\
transitions f$_{1}$ (=17.2 T) and f$_{2}$ (=25 T) for the frequency of 1000 GHz (=48 K) are shown - they are in perfect
agreement with recent experimental observation of Ref. \cite{42}. The zero energy is at the level of the unsplit
$^{5}D$ term. The ground term is subterm $^{1}$A$_{1}$ originating from the $^{1}$I term that is shifted down by 4.43
eV by octahedral crystal-field interactions.}
\end{center}
\end{figure}
Recently it has been unambiguously proved by very sophisticated
electron-spin-resonance experiments on single crystal \cite{42} in
external magnetic field up to 30 T and the frequency up to 1390
GHz ($\hbar \omega$ = 66.7 K = 5.75 meV) that such the
singlet-doublet structure, as shown in Fig. 3, with a splitting of
0.6 meV only is realized in LaCoO$_{3}$ indeed, Fig. 11
\cite{18,42}, but there is about 12 meV lower another singlet as
the ground state. This singlet ground state is a $^{1}A_{1}$
subterm and originates from the $^{1}I$ term (S=0, L=6). The
$^{1}I$ term in the free Co$^{3+}$ ion lies 4.43 eV above the
$^{5}$D term \cite{21}, but in LaCoO$_{3}$ in the presence of the
octahedral crystal field interactions the $^{1}I$ term is split
into six subterms and a subterm $^{1}$A$_{1}$ ($^{1}I$) lowers so
much its energy due to the very large orbital moment of the
$^{1}I$ term (L=6) and substantial octahedral crystal-field
interactions, Fig. 12. The octahedral CEF turns out to be about
25\% stronger than we originally thought \cite {43,44,45} (instead
of $B_{4}$ of 200 K it turns out to be of 260 K) \cite{18}, but
these octahedral CEF interactions are still not so strong to break
intra-atomic correlations among electrons within 3d shell (the
preservation of dominant intra-atomic correlations among electrons
within 3d shell, leading to the 3d$^{6}$ configuration, is our
meaning of ''atomistic'').

In Fig. 11 we present our calculations of the behavior of the
three lowest Co$^{3+}$ states of the fine electronic structure
shown in Fig. 3 under the action of the external field up to 70 T.
The calculated results for the energy separations perfectly
reproduce the experimental data measured up to 30 T for the field
applied along the cube diagonal as well as for other two main
crystallographic directions. A remarkably good description of the
experimentally derived quasi-triplet states with its behavior in
magnetic fields up to 30 T applied along different main
crystallographic directions \cite{18} proves the high physical
adequacy of the used by us intermediate CEF approach to 3d-ion
compounds in contrary to the generally used strong CEF approach.
Despite of a non-Hund's rule ground state of the Co$^{3+}$ ion in
LaCoO$_{3}$ the QUASST theory is still valid for LaCoO$_{3}$ as
the $^{1}A_{1}$ subterm is the term expected from the atomic
physics. In fact, we never expected that in a 3d solid electronic
states will be so thin, in the energy scale below 1 meV, and so
well characterized by the atomic physics. The recent ESR results
on LaCoO$_{3}$, its theoretical anticipation and the perfect
description provide strong, if not unambiguous, evidence for the
existence of the discrete atomic-like fine electronic structure in
the 3d-atom containing compound confirming the basic concept of
QUASST.
\begin{figure}[!ht]
\begin{center}
\includegraphics [width=11.5 cm , bb= 20 20 573 772]{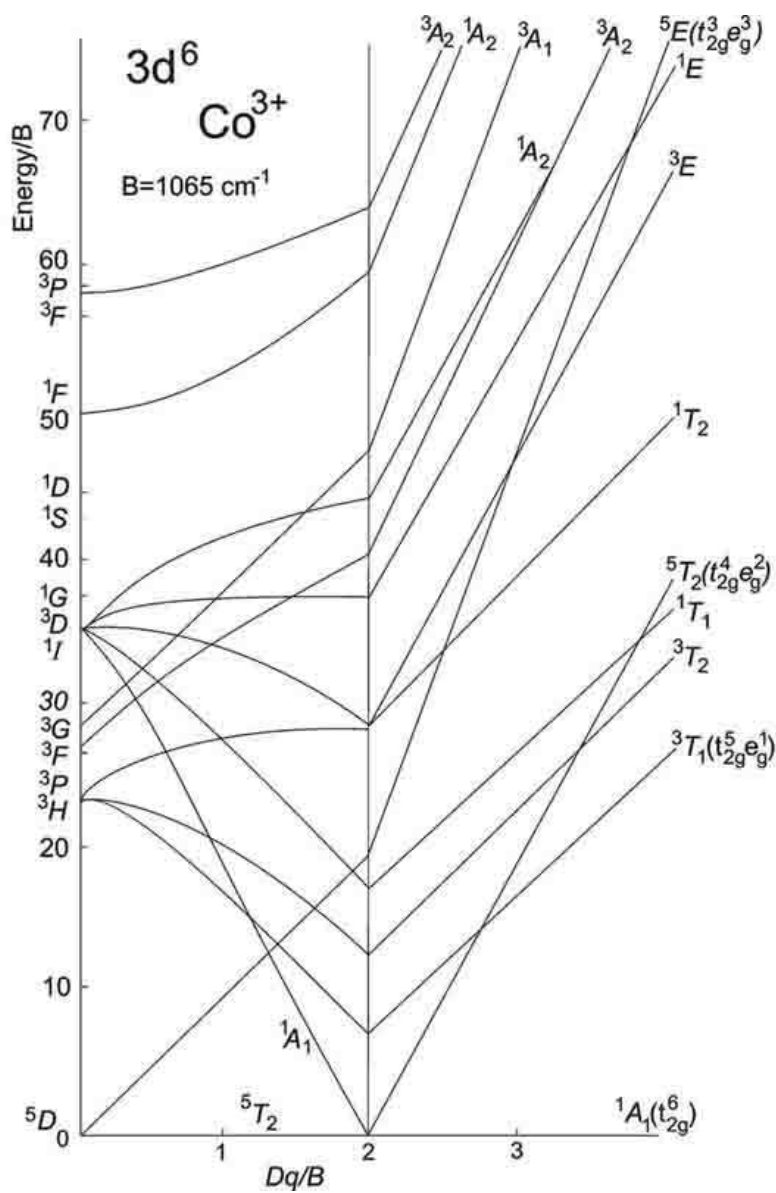}
\caption{\footnotesize{Influence of the strength of the octahedral crystal field interactions on the atomic terms of
the 3d$^{6}$ configuration occurring in Co$^{3+}$ and Fe$^{2+}$ ions. After Tanabe and Sugano \cite{31}. The zero
energy is at the level of the unsplit $^{5}$D term. At strong crystal field subterm $^{1}$A$_{1}$, originating from the
$^{1}$I term, is shifted down by 4.43 eV by octahedral crystal-field interactions becoming the ground term instead of
Hund's rules $^{5}$D ground term. In LaCoO$_{3}$ Dq/B, Dq=12B$_{4}$ and B is Raccah parameter B=1065 cm$^{-1}$ for the
Co$^{3+}$ ion, amounts to 2.05 and the $^{1}$A$_{1}$ term is only 140 K below the $^{5}$D. The value Dq/B=2.05 in
LaCoO$_{3}$ corresponds to B$_{4}$=+260 K. For the Fe$^{2+}$ ion in FeBr$_{2}$ Dq/B$\simeq$1.4.}}
\end{center}
\end{figure}

The QUASST predicts next ESR absorptions at the field of 40-42 T
for the 1000 GHz frequency, Fig. 11, provided non-zero matrix
elements. There is also expected at a field of 60 T a level
crossing and the transition to the magnetic state. It will be the
field induced magnetic state.

The QUASST description proves that in LaCoO$_{3}$ there is no IS
state as comes out from band-structure calculations \cite{9}.
Moreover, the nature of the HS state is completely different from
that shown in Fig. 2c. According to Fig. 2c the HS state,
characterized by S=2, is a spin-polarized state, caused by
J$_{ex}$, what is wrong owing to the fact that in LaCoO$_{3}$,
being not magnetically-ordered, there is no internal magnetic
field and no spin polarization. Our HS state, also with S=2, is
not-spin-polarized state in accordance with the
experimentally-observed paramagnetic state.
\begin{figure}[!ht]
\begin{center}
\includegraphics [width=10 cm , bb= 20 20 592 427]{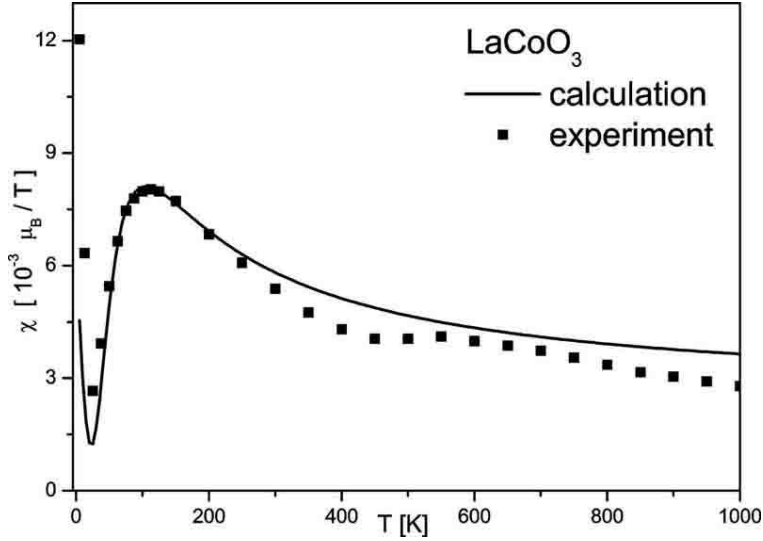}
\caption{Calculated temperature dependence of the paramagnetic susceptibility $\chi$(T) of LaCoO$_{3}$. Calculations
have been performed for the fine electronic structure as shown in Fig. 3, but taking only into account one third of Co
atoms and a diamagnetic contribution of -1.5$\cdot$10$^{-3}$ $\mu_{B} $/T (= -0.83$\cdot$10$^{-3}$ emu/mol).
Low-temperature upturn is due to impurities, about 1 $\%$ of the Co$^{3+}$ ions in the $^{5}$D term ground state.
Points denotes recent experimental values, rough data of Ref. \cite{46}.}
\end{center}
\end{figure}

There is a problem left whether the electronic triplet structure
derived from ESR is representative for {\bf all} Co ions in
LaCoO$_{3}$ or may be the observed ESR signals originate from some
impurity atoms. Impurity atoms can be also Co ions on the surface
and others not being full part of the crystallographic structure.
In Fig. 13 the temperature dependence of $\chi_{d}$(T) calculated
for the electronic structure shown in Fig. 3 is presented - the
calculations reproduce the overall shape of the temperature
dependence in the whole temperature range but the agreement with
the experimental maximal value at 100 K has been reached by
multiplying by N$_{A}$/3 only - it could mean that only one third
of Co ions contribute to the observed susceptibility. In
calculations we took into account a diamagnetic contribution of
-1.5$\cdot$10$^{-3}$ $\mu_{B} $/T (= -0.83$\cdot$10$^{-3}$
emu/mol) according to formula
\begin{equation}
\chi_{exp} \\=\\ \chi_{d} + \chi_{dia}
\end{equation}
showing that the expected experimental value should be lower than
the calculated value $\chi_{d}$. In literature the opposite
treatment of the diamagnetic contribution is made for LaCoO$_{3}$,
see Ref. \cite{46}, for instance, but we think that our treatment
is correct.

For the explanation of this discrepancy, too large calculated
susceptibility, we would recall many reasons like the covalency
(but somebody has to prove it clearly defining its effect on
physical properties, in particular on the susceptibility), the
hybridization (the same as above), the reality is more complicated
than the single-ion aproach and so on. We are looking for for more
physical and better-defined explanations, that we require, and we
think that the smaller observed susceptibility can be caused by a
zig-zag arrangement of the oxygen octahedra in the distorted
perovskite structure (Co-O-Co angle in LaCoO$_{3}$ is not
180$^{\circ}$ but amounts to 163$^{\circ}$ only \cite{16}) but we
would suggest that indeed only 1/3 of Co ions contribute to the
susceptibility observed at 100 K. It would require that local
distortions distinguish Co sites in a proportion 1 to 2 and that
in the second sites the octahedral crystal field is larger
yielding $\Delta$$_{s-t}$, say, of 1000 K. It would cause that
their contribution to $\chi$(T) would be experimentally invisible.
For such the increase of B$_{4}$ the tiny closer oxygen-cobalt
distance, less than 1 percent, is needed and surely cannot be
excluded from the experimental point of view.

The calculated results, taking into account one third of Co ions,
well describe the recent experimental result for $\chi(T)$
\cite{46} as is seen in Fig. 13. For detailed comparison one
should remember that $\Delta$$_{s-t}$ in LaCoO$_{3}$ is very small
and that there are going very subtle effects with temperature.
With temperature due to conventional thermal expansion the
strength of octahedral CEF interactions slightly decreases. This
thermal expansion effect is small but in case of LaCoO$_{3}$ it is
enough to reverse the ground state. Taking into account the
thermal expansion result of Ref. \cite{47} we have calculated that
$\Delta$$_{s-t}$ becomes zero at T about 500-600 K - we think that
this level crossing is responsible for the anomaly in the
$\chi$(T) dependence \cite {7,46} in this temperature region. The
effect opposite to the thermal expansion, i.e. the decrease of the
Co-O distance and the increase of the B$_{4}$ parameter can be
produced by external pressure. Then, with the applied pressure the
maximum temperature of $\chi$(T) is shifted to higher temperatures
indicating the increase of $\Delta$$_{s-t}$ \cite{48,49}
consistently with the increase of $B_{4}$.

All of these facts show that LaCoO$_{3}$ is very unique system,
indeed. Thanks physical coincidences the ground state is close to
the level crossing of states related to different terms. It
enables exact evaluation of the strength of CEF interactions and
to study the Stark effect on different terms.

\subsection{Example III: Nonmagnetic Mott insulator: Na$_{2}$V$_{3}$O$_{7}$}

In 2002 Gavilano {\it et al.} \cite{50} have discovered a drastic
violation of the Curie law in Na$_{2}$V$_{3}$O$_{7}$ in the
experimentally measured temperature dependence of the magnetic
susceptibility. Analysis of the $\chi$(T) dependence in terms of
the Curie law shows that the effective moment of the V$^{4+}$ ion
is reduced by the one order of magnitude upon reducing the
temperature from 100 to 10 K. Gavilano {\it et al.} have provided
an explanation that ''The reduction of the effective magnetic
moment is most likely due to a gradual process of moment
compensation via the formation of singlet spin configurations with
most but not all of the ions taking part in this process. This may
be the result of antiferromagnetic interactions and geometrical
frustration.'' Using some crystallographic-arrangement arguments
they further conjectured ''the compensation of eight out of the
nine V spins ...'' in order to reproduce the observed reduction of
the effective moment by one order of magnitude. There is no sign
of the magnetic order down to 1.9 K \cite{50}. So we have to
accept that Na$_{2}$V$_{3}$O$_{7}$ is in the paramagnetic state
down to this temperature.

We could reproduce very well the observed temperature dependence
of the paramagnetic susceptibility by considering the electronic
structure associated with the V$^{4+}$ ion (3d$^{1}$
configuration) under the action of the crystal field taking into
account the spin-orbit coupling. Despite of the Kramers doublet
ground state, a ground-state with quite small magnetic moment is
obtained as an effect of the spin-orbit coupling. It turns out
that even weak s-o coupling unquenches a quite large orbital
moment. In our explanation we have been oriented by our earlier
calculations for the V$^{4+}$ ion presented in Refs \cite{51}.

\begin{figure}[!ht]
\begin{center}
\includegraphics [width=10.8 cm , bb= 20 20 534 772]{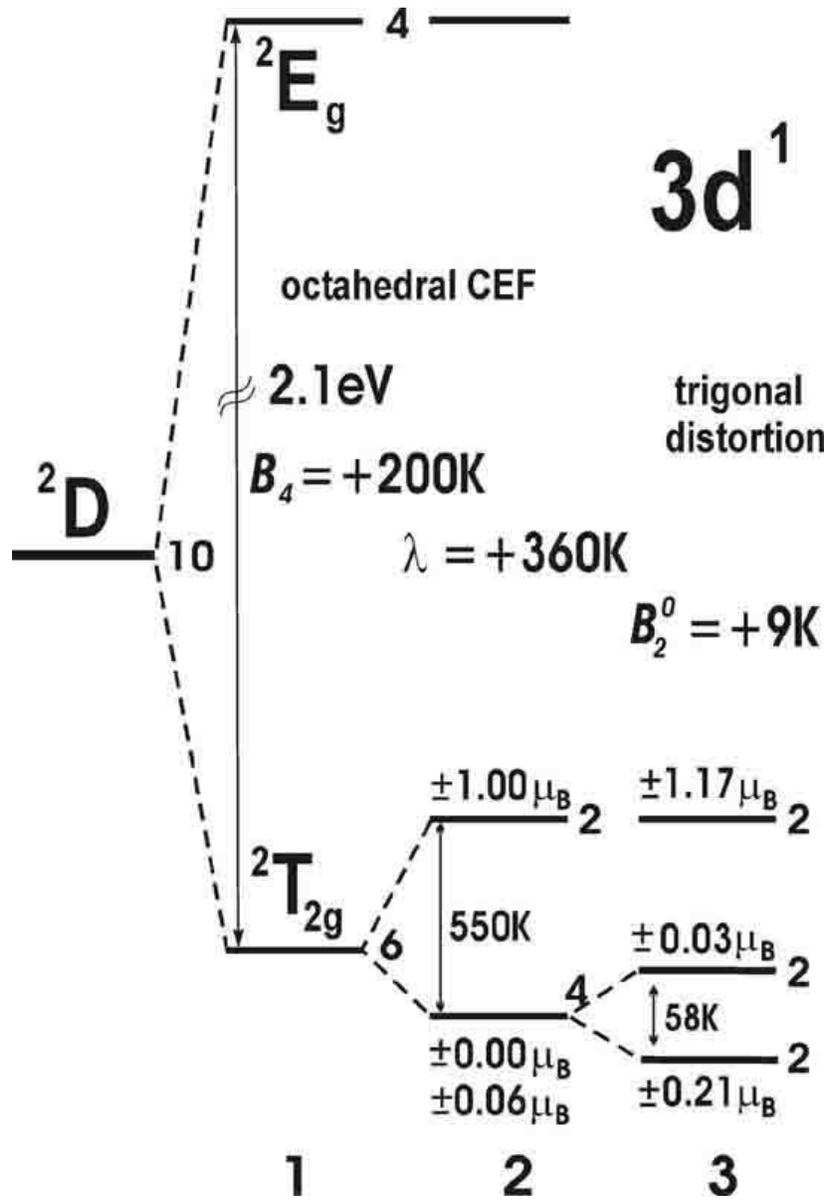}
\caption{Fine electronic structure of the strongly-correlated 3d$^1$ electronic system related to the 10-fold
degenerated Hund's rules $^2$D ground term. 1) the effect of the oxygen octahedral crystal field, 2) the effect in
combination with the spin-orbit coupling; 3) further splittings due to trigonal distortions. The calculated resulting
susceptibility is shown in Fig. 15 and it well reproduces experimental results for Na$_{2}$V$_{3}$O$_{7}$.}
\end{center}
\end{figure}

The one 3d electron in the V$^{4+}$ ion is described by quantum
numbers $L$=2 and $S$=1/2. The ground term $^{2}D$ is 10-fold
degenerated. Its degeneracy is removed by the intra-atomic
spin-orbit interactions and in a solid by crystal-field
interactions. This situation can be exactly traced by the
consideration of a single-ion-like Hamiltonian \cite{51,52}
\begin{equation}
H_{d}=H_{CF}^{octa}+H_{s-o}+H_{CF}^{tr}+H_{Z}=B_{4}(O_{4}^{0}+5O_{4}^{4})+\lambda
\mathbf{L}\cdot \mathbf{S}+B_{2}^{0}O_{2}^{0}+\mu
_{B}(\mathbf{L}+g_{e}\mathbf{S})\cdot \mathbf{B}
\end{equation}

in the 10-fold degenerated spin-orbital space. We approximate, for
simplicity, the CEF interactions at the V site by considering
dominant octahedral interactions with a trigonal distortion. For
the octahedral crystal field we take $B_{4}$= +200 K. The sign
''+'' comes up from {\it ab initio} calculations for the ligand
octahedron. The spin-orbit coupling parameter $\lambda _{s-o}$ we
take as +360 K, as in the free V$^{4+}$ ion \cite{27}, p. 399.

The resulting electronic structure of the 3d$^{1}$ ion contains 5
Kramers doublets separated in case of the dominant octahedral CEF
interactions into 3 lower doublets, the $T_{2g}$ cubic subterm,
and 2 doublets, the $E_{g}$ subterm, about 2 eV above (Fig. 14).
The $T_{2g}$ subterm in the presence of the spin-orbit coupling is
further split into a lower quartet and an excited doublet (Fig.
14(2)). Positive values of the trigonal distortion parameter
$B_{2}^{0}$ yields the ground state that has a small magnetic
moment (Fig. 14(3)). For $B_{2}^{0}$ = +9 K the ground state
moment amounts to $\pm 0.21$ $\mu _{B}$. It is composed from the
spin moment of $\pm 0.48$ $\mu _{B}$ and the orbital moment of
$\mp 0.27$ $\mu _{B}$ (antiparallel). The sign $\pm $ corresponds
to 2 Kramers conjugate states. The excited Kramers doublet lies at
58 K (5 meV) and is almost non-magnetic - its moment amounts to
$\pm 0.03$ $\mu _{B}$ only (=$\pm 1.03$ $\mu _{B}+2\cdot (\mp
0.50$ $\mu _{B})$) due to the cancelation of the spin moment by
the orbital moment. So small and so different moments for the
subsequent energy levels is an effect of the spin-orbit coupling
and distortions.

\begin{figure}[!ht]
\begin{center}
\includegraphics [width=11.3 cm , bb= 20 20 592 437]{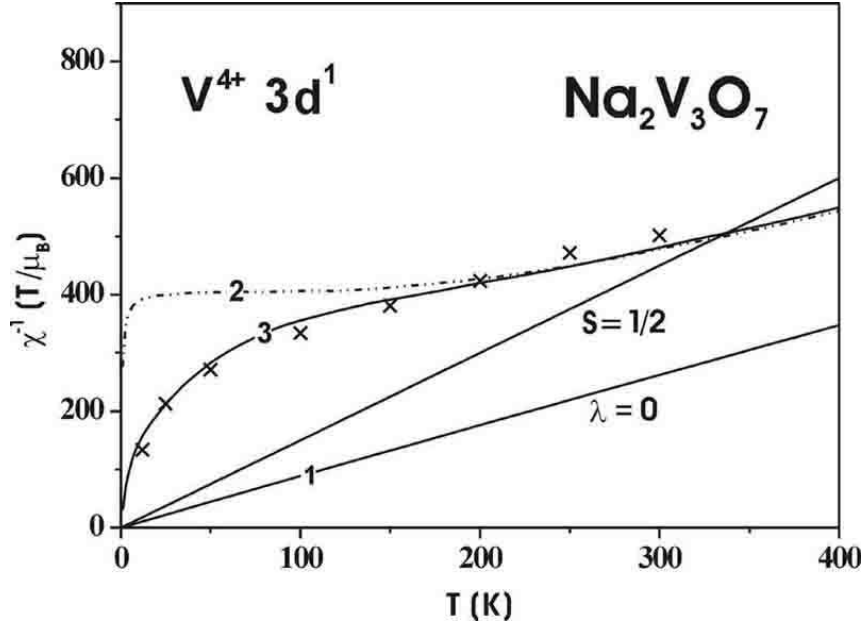}
\caption{Temperature dependence of the paramagnetic susceptibility $\chi$(T) of Na$_{2}$V$_{3}$O$_7$ (curve 3)
calculated for B$_4$=+200 K, $\lambda$$_{s-o}$ = +360 K and the trigonal parameter B$_{2}^{0}$ = +9 K. Curve 1 -
results for the octahedral crystal field without the spin-orbit coupling. Curve 2 - the octahedral crystal field with
the spin-orbit coupling \cite{52}. Curve marked as S=1/2 shows $\chi$(T) expected from Curie law with S=1/2. Crosses
are rough experimental values of Ref. \cite {50}.}
\end{center}
\end{figure}

The resulting susceptibility is shown in Fig. 15. It is clear that
we could describe overall temperature dependence of $\chi$(T)
\cite{52}. For comparison of the calculated susceptibility value
with experimental data we follow Eqn. 9 taking into account a
diamagnetic contribution of -0.7$\cdot$10$^{-3}$ $\mu_{B} $/T
V-ion (= -0.4$\cdot$10$^{-3}$ emu/mol V).

We are fully convinced that the observed violation of the
Curie-Weiss law in Na$_{2}$V$_{3}$O$_7$ is the effect of the
spin-orbit coupling and CEF interactions. From almost perfect
reproduction of the absolute value we can say that all V ions are
in the tetravalent state and that all V$^{4+}$ ions contribute
equally to the susceptibility.

QUASST predicts Na$_{2}$V$_{3}$O$_7$ to exhibit a Schottky-like
peak at 25 K and below 1 K a large low-temperature specific heat,
resembling heavy-fermion phenomena, as the Kramers degeneracy of
the ground state has to be removed before reaching the absolute
zero temperature. Obviously the study of Na$_{2}$V$_{3}$O$_7$ at
low temperatures is of large scientific interest - a magnetic
ordering at ultra-low temperatures is discussed recently in terms
of the quantum critical point.

\section{Further remarks about QUASST in a solid}
\subsection{Single ion \emph{vs} a three dimensional solid}

The crystal-field approach itself is usually considered as the
single-ion approach and then arguments are given against its
usefulness in the discussion of a real three dimensional (3D)
solid. We disagree with such a thinking and in order to point out
the physical relevance of the CEF concept to a solid the more
general name, Quantum Atomistic Solid State Theory (QUASST), has
been given. At first, it is important to notice that we consider
in QUASST not a single 3d cation but the cation in the octahedral
surroundings, i.e. the complex MO$_{6}$. In the crystal there is
the translational symmetry and the whole 3D crystal of, say NiO
with the NaCl structure, is built up from the face-sharing
octahedra along three main crystallographic directions.

The CEF parameters reflect interactions of the charge {\bf
multipolar} moment of the 3d cation with the charge multipolar
moments of the surroundings of the {\bf whole} crystal, not only
of the local octahedron.

\subsection{Jahn-Teller effect - fundamental importance of the
spin-orbital space}

The Jahn-Teller (J-T) effect can be easily illustrated in the
electronic structures presented in Figs 5c and 6c. According to
the Jahn-Teller theorem there is always a spontaneous distortion
of the local lattice surroundings going on in order to remove the
degeneracy of the ground state. Inspecting Figs 5c and 6c we can
say that lattice off-cubic distortions will proceed for all 3d
cations in order to remove the ground state degeneracy. The
subsequent manifestation of the J-T effect is to increase the
energy separations among the lifted-degeneracy levels. The quartet
degeneracy of the lowest state in case of the d$^{1}$
configurations can be lifted only to two doublets owing to the
Kramers theorem as is calculated in Fig. 14. Similarly in case of
the d$^{3}$ and d$^{9}$ configurations. The splitting of the
triplet in case of the 3d$^{6}$ configuration into (lower)singlet
and excited doublet has been analyzed already in case of
LaCoO$_{3}$ despite the fact that this triplet is the excited
state. It reveals more subtle Jahn-Teller effect on the excited
states. In FeBr$_{2}$ the triplet degeneracy is removed, see Fig.
8, yielding the lower doublet and higher singlet structure. Such
the structure is somehow in disagreement with the subtle
Jahn-Teller effect - Nature reaches, however, the singlet ground
state in FeBr$_{2}$ by the formation of the magnetic state, Fig.
9a. This discussion can be concluded that all 3d ions are
practically Jahn-Teller ions, i.e. that in all compounds we should
expect the local off-octahedral distortions to occur.

One should note that the present discussion of {\bf the
Jahn-Teller theorem in the spin-orbital space} differs completely
from the removal of the degeneracy in the orbital-space only as is
customarily discussed in the present literature related to iron
group compounds. It is obvious that the definition of the physical
space is crucial for the proper analysis of the Jahn-Teller
theorem. We claim that the spin-orbital space is the physically
proper space for 3d-ion compounds \cite{53}. For rare-earth
systems the Jahn-Teller effect is discussed using the total
angular momentum quantum number J as the good quantum number. Such
the J space of the three Hund's rules ground multiplet is
equivalent to the spin-orbital space, because J becomes the good
quantum number as a result of the strong spin-orbit coupling, in
fact infinitively strong. It means, that thanks taking into
account the spin-orbit coupling in description of 3d ions we make
the unification in description of the electronic structures of 3d
and 4f ions and compounds.

\subsection{Symmetry breaking with lowering temperature as a general QUASST
rule}

In the QUASST we point out that with decreasing temperature there
appears more and more breaking symmetry mechanisms in order to
remove the ground-state degeneracy and to increase the energy
separation to the excited states. Here we can mention a lowering
of the local symmetry, the site differentiation and the formation
of magnetic state. The magnetic state related to the time reversal
symmetry breaking often appears with the lowering temperature. In
case of the site differentiation instead of the single energy
level scheme for the crystal (what means, the same energy scheme
at each 3d site) we can have many energy schemes superimposed on
each other originating from different sites. It could resemble a
band. Of course, the nature of this band is completely different
from the band in the band theories. This band-like closely lying
CEF states are due to the site differentiation and a structural
disorder. One should note that for the band formation the perfect
translational symmetry is needed. Our atomistic approach can be
used for a low-symmetry, non-stoichiometric systems and even for
amorphous systems.

LaCoO$_{3}$ undergoes the rhombohedral distortion at 1610 K
\cite{54}. A low-tempera\-ture upturn below 30 K in $\chi$(T), see
Fig. 13, is supposed to be due to about 1 percent impurities or Co
atoms/ions on the surface, i.e. atoms with largely reduced
dimensionality and very low symmetry.

\subsection{Zeeman effect and evaluation of the internal molecular field}

An effect of the internal or external magnetic field on the
electronic states, known as Zeeman effect, can be nicely analyzed
on states shown in Figs 5c and 6c with the use of Hamiltonian (3).
In the magnetically-ordered state in a solid there appears the
internal magnetic field. Its effect on the lowest triplet of the
Fe$^{2+}$ ion in FeBr$_{2}$ is shown in Fig. 9a. We have
calculated that in FeBr$_{2}$ at 0 K the Fe$^{2+}$-ion moment
experiences the molecular field B$_{M}$(0) of 5.15 T \cite{17}.
This field diminishes with temperature becoming zero at T$_{N}$ =
14.2 K. In LaMnO$_{3}$, an antiferromagnet with T$_{N}$ of 140 K,
B$_{M}$(0) = 108 T \cite{55}. In CoO, an antiferromagnet with
T$_{N}$ of 290 K, B$_{M}$(0) = 169 T \cite{41}. In NiO, an
antiferromagnet with T$_{N}$ of 525 K, B$_{M}$(0) = 500 T
\cite{38}. In general, the higher T$_{c}$(T$_{N}$) the larger
molecular field is.

\subsection{Clear distinction between magnetically-ordered and paramagnetic state}

In QUASST there is clear distinction between the
magnetically-ordered and the paramagnetic state. In the magnetic
state the time reversal symmetry (TRS) is broken. This TRS
breaking is possible to trace in the atomic scale as then appears
the magnetic spin-like polarization of the eigenfunctions and
states. The Kramers doublets become split and then there appears a
spin-like gap of a few meV size, depending on values of T$_{c}$
(T$_{N}$). In the magnetic state the mean value of the operator
$\langle$$\hat{J}$$_{z}$$\rangle$ in case of rare-earth ions and
the mean values of $\langle$$\hat{S}$$_{z}$$\rangle$ and
$\langle$$\hat{L}$$_{z}$$\rangle$ operators defining the total
magnetic moment by ($\hat{L}$$_{z}$+g$_{e}$$\hat{S}$$_{z}$), in
case of iron-group ions, becomes different from zero. These values
of the ground state do not have any direct connection to the
effective moment derived from the paramagnetic susceptibility,
that is a measure of $\hat{J}$$^{2}$ and $\hat{S}$$^{2}$
operators. In the paramagnetic state the time-reversal symmetry is
preserved despite that the local moment is still described by
quantum numbers S$\neq$0 and $L\neq$0.

In QUASST the paramagnetic-magnetic transition is nicely describes
in the temperature dependence of the heat capacity c(T), as it was
shown in Fig. 9c for FeBr$_{2}$. It is worth noting that in QUASST
the overall c(T) dependence is reproduced both in the absolute
value and the $\lambda$-type peak at T$_{c}$ (T$_{N}$). It is a
surprise that already so simple mean-field Hamiltonian, the third
term in Eqn (7), provides so good reproduction of the
$\lambda$-type peak.

\subsection{Single-ion magnetism \emph{vs} collective nature of the
magnetic state}

One can think that the presented approach is too simple. Surely it
is, but the most important is that it is confirmed by experiment.
So, the physics of solids could be complex, but it turns out that
Nature is not so complicated as could be. God is gracious to us!
However, the evaluation of CEF interactions, effects of the
lowering symmetry with decreasing temperature, the site-to-site
change of the quantum local z-axis (so-called zig-zag alignment of
octahedra in real perovskite compounds) and a possible site
differentiation with respect to the charge and local surroundings
in cooperation with the spin-orbit coupling and magnetic
interactions make the problem far from being easy to tackle for
real compounds. It is the reason that though the concept of the
crystal field is known already by 75 years there is a quite
limited number of compounds in which CEF interactions and magnetic
interactions have been unambiguously evaluated. In particular, for
3d-atom containing compounds. In pointing out, in our presentation
of QUASST, of the importance of the atomistic description we would
like to remind an obvious fact that it is atoms and ions that are
the source of magnetism of solids.

\subsection{Many-electron \emph{vs} single-electron states}

In QUASST we deal with many electron states and consequently with
many-electron wavefunctions in contrary to single-electron states
and single-electron wavefunctions considered in the band theories.
In general, the wavefunction and the energy of the whole
atomic-like 3d$^{n}$ many-electron system of a given cation can be
constructed from single-electron wavefunctions. In QUASST it is in
fact assumed that this many-electron construction from
single-electron wavefunctions will lead to the atomic-like
wavefunctions. Other words, we think that the ambitious \emph{ab
initio} theories starting the description of a solid from a
(gaseous) jellium of independent electrons have to put such
conditions and introduce such mutual electron interactions into
their theories in order to get atoms or given ions. In QUASST we
take the existence of atoms or ions as the physical fact that is
not necessary to prove. Their existence in the reality confirms
the validity of our assumption. We treat the problem of
description of atoms or ions as the subject of atomic physics.

\subsection{The nature of the states}

The states discussed in QUASST are physical states - they can be
experimentally observed by energy spectroscopic methods, like the
inelastic neutron scattering, the light absorption, or the
Electron Spin Resonance. The existence of these states give
physical effects like an extra heat capacity, quite pronounced for
low-energy states giving a non-trivial Schottky-like heat
contribution at low temperatures, or a violation of the
Curie-Weiss law at low temperatures. We point out the physical
nature of the discussed states as in presently-in-fashion theories
often virtual states and energy transitions of 2-5 eV are
considered to give substantial physical effects below the room
temperature.

In QUASST the number of involved states is counted by the
configurational entropy. This configurational entropy is
experimentally measured by the integration of the heat capacity,
after the substraction of the lattice heat. The presence of a
large number of electronic states in energies below 25 meV is made
use in the practice for magnetocaloric materials. The related
entropy is freed at the magnetic ordering temperature. Inspecting
Figs 5 and 6 we expect large magnetocaloric effect to occur for 3d
ions in the octahedral voids for the Mn$^{3+}$ ions - there
entropy related to 10 lowest states can be freed.

\subsection{Insulating ground state}

Within the QUASST the insulating ground states of oxides is easily
understood. After the charge transfer during the formation of a
compound later electrons are bound to atoms. They are staying
localized in case of the perfect stoichiometry and the perfect
crystallographic structure. In oxides the conductivity appears as
an effect of an imperfection of the crystal structure,
off-stoichiometry, the appearance of vacancies and, in general, as
disorder effects. There can appear the conductivity in perfect
crystals in case of, for instance, the existence of few sites with
ions with different valences. This latter case occurs in
Fe$_{3}$O$_{4}$, in which Fe$^{2+}$ and Fe$^{3+}$ ions coexist at
different crystal sites. Thus, the conductivity in oxides is in
general of the hopping mechanism and has the activation character.

In LaCoO$_{3}$ the resistivity dramatically grows up with
decreasing temperature reaching at 60 K already 10$^{7}$
$\Omega$cm \cite{4} confirming that LaCoO$_{3}$ is the Mott
insulator despite of the existence of the incomplete 3d shell.
With increasing temperature the resistivity smoothly drops
reaching at about 700 K a value of 10$^{-3}$ $\Omega$cm which is a
Mott border-resistivity value for beginning of the metallicity.
Some people discuss this 10 order of magnitude reduction of the
resistivity as an insulator-metal transition, though there only
smooth crossover occurs with temperature.

In intermetallic compounds there are itinerant electrons, apart of
bound localized electrons, as it was discussed for ErNi$_{5}$ and
NdNi$_{5}$ \cite{21,25}.

\subsection{Pauli exclusion principle and Boltzmann statistics}

For description of thermodynamical properties in CEF and QUASST
theories the Boltzmann distribution function is used. The Pauli
exclusion principle is made use in the description of particular
atoms, i.e. in the description of electrons in shells, that
appears in the allowed terms. The Boltzmann distribution function
is used for the calculation of thermodynamical properties because
it describes the thermal equilibrium of the collection of atoms
having available different energy states.

\subsection{Unification of description of 3d and 4f paramagnetic
ions}

Rare-earth ions (4f) are customarily described using J as the good
quantum number \cite{21,24,25}. This the one-multiplet
approximation gives physically adequate result because the
spin-orbit coupling is large. Due to the strong s-o coupling the
ground multiplet is well separated from the exited multiplets.
Thus, the description within the ground multiplet, resulting from
three Hund's rules, is a justified approximation of calculations
within the $|$LSL$_{z}$S$_{z}\rangle$ space. In contrast, 3d
paramagnetic ions are customarily described, if not a one-electron
approach is used, in the separate L and S spaces. It means, that
the magnetic moment is the spin-only moment, whereas the
electronic structure is built up from orbital states only. In our
approach we describe both 4f and 3d paramagnetic ions within the
same spin-orbital space $|$LSL$_{z}$S$_{z}\rangle$. Thus, QUASST
provides the unified description of 3d and 4f paramagnetic ions.

\section{Theoretical background of the crystal field and \\QUASST}

\subsection{Old arguments against the crystal-field approach}
As mentioned already, after very successful achievements of Van
Vleck and his associates in years of 1929-1953 \cite{56,57,58,59}
Slater in 1953 \cite{60} came out with, according to Stevens in a
paper written in 1979 \cite{61}, p. 4, a strong attack on the
crystal field theory. The attack must have been really very strong
as many researchers even very successful in the application of the
CEF theory to real systems drop the crystal-field ideas retaining,
as Stevens wrote \cite{62}, the spin-Hamiltonian as an effective
description only. The main objection of Slater was that the
crystal-field theory violates the physical principle of the
indistinguishability of electrons. Presenting Slater's argument
Stevens wrote in Ref. \cite{61}, p. 5, that by fact that in the
CEF theory we treat a single magnetic ion in a crystal lattice,
say the Nd$^{3+}$ ion for example, the electrons become
distinguished by ascribing electrons 1, 2 and 3 to the 4f shell
and regarding all the others as simply constituting part of the
electrostatic field. According to us this argument is not
appropriate - \textbf{in the CEF theory they are states but not
electrons that become distinguished.} Other argument of Slater,
also related to the electron indistinguishity principle, was that
electrons are distinguished by ascribing them to the respective
sites. Again the Slater's understanding of the electron
indistinguishity principle is erroneous in case of solids - this
ascribing of electrons is to the atoms and the electrons are
localized being bound to its original atom. One could say,
remembering a term of slave bozons popular in modern solid-state
theory, that electrons in CEF and in QUASST are slaves of atoms.

Later questioning of CEF in seventies of XX century was by Freeman
\cite{63}, pp 321-327, for the inability to calculate exactly CEF
parameters. There was no doubt that author of Ref. \cite{63}
favors the band theory. We think that the inability argument
should be taken as critics of a detail of the CEF theory, like
point-charge approximation, rather than the CEF theory itself.
According to Freeman other environmental effects, overlap and
covalency should have much bigger effect on the crystal-field
splitting than the electrostatic field description. We could only
very partly agree with this point but simultaneously we would like
to point out that a Freeman's 30-year-old hope that the band
theory {\it{ab initio}} soon solve the description of magnetic and
electronic properties still awaits for the realization. As far as
exact calculations of CEF parameters are discussed we would like
to say that all CEF interactions, in particular for rare-earth
compounds that have been of the Freeman interest, are of very
small energies, say of 10 meV. It is at least 2-3 orders of
magnitude weaker than potentials discussed in the band theory.
Thus, we prefer to reverse this scientific problem - the
evaluation of the low-energy electronic structure and its
parametrization by means of \textbf{CEF parameters can be used for
the evaluation of the charge distribution in the elementary cell}.
In such the methodology we use, in fact, 3d/4f/5f cations as a
microscopic probe for the existence of the crystal-field
multipolar potentials in a solid. Such the evaluation is only
possible provided the full energy level scheme is perfectly
established. The full scheme, not only the ground state, is needed
for the meaningful evaluation of different multipolar
contributions. Moreover, we would like to point out that the CEF
theory and QUASST reveal the need of 0.5 meV accuracy in the
evaluation of the electronic structure. It is 1000 times more
exact than the presently attained accuracy of band-structure
calculations.

The questioning of the crystal-field theory has caused that today,
despite of the fact, that there is continuously growing number of
compounds described in terms of the CEF Hamiltonian the
crystal-field theory itself, often identified with the shortage of
the point-charge model, is often regarded as "naive". We fully
disagree with a statement of Freeman \cite{63}, p. 322, that "The
success of this naive (and from the theoretical point of view
inadequate) model is thought to lie in the fact that V$_{c}$ gives
the Hamiltonian the correct symmetry and a sufficient number of
empirical parameters which somehow absorb all the various
environmental effects not included in the simple description given
by Eqn (6.49)", i.e. by a CEF potential expanded in terms of
spherical functions.

Other aspect of the problem of a disdainful treatment of the
crystal field is related to the education - students of physics
are not taught, or only very little, of the crystal field and of
the group theory.

\subsection{Theoretical basis for the crystal-field and QUASST}

A basic message of the QUASST is that the low-energy discrete
electronic structure of the atomic-like 3d$^{n}$ configuration,
determined by crystal-field and spin-orbit interactions, is
largely preserved in a solid 3d-atom containing compound.
According to us \textbf{the theoretical basis for the
crystal-field theory is the atomistic construction of matter} and
we are very surprised that it was not recognized earlier. The
atomistic construction of matter is the basic well-established
principle of Nature - one could say that this knowledge is already
known from 2 500 years, thinking about loosely ideas of
Democritos, another one, that from 100 years, when atoms have been
accepted in physics to exist in the reality (1913 - Bohr's model
for the hydrogen). Thus we are fully convinced that the crystal
field and QUASST theory are based on well-known and
well-established physical concepts, like the local symmetry, the
group theory, atomic physics with Stark and Zeeman effects,
statistical physics and many others. The crystal field and QUASST
theory discuss the discrete states, the existence of which is a
characteristic feature of the quantum mechanics.

In such the formulation QUASST can be experimentally verified and,
as Carl Popper says in the scientific methodology, it can be
falsificated. The falsification possibility is the basic
requirement, according to Popper, demanded for a scientific
theory. We are ready for the open scientific discussion about the
physically adequate description of the electronic structure and
the magnetism of 3d-/4f-/5f-atom containing compounds.

We are convinced that the CEF theory not only provides a
conceptually simple framework for the interpretation of numerous
experimental results, as many people are ready to accept the
crystal field approach, but the CEF theory must be a basic
ingredient of each modern solid state theory dealing with
electronic and magnetic properties of transition-metal compounds.

QUASST reconciles the CEF theory with localized moments with the
lattice periodicity, a problem, that according to Stevens
\cite{62}, p. 7, has divided magneticians for many years.

In QUASST d electrons take part both in the bonding and in the
magnetism. The primary bonding proceeds via electrostatic
interactions. The magnetic interactions, in fact spin-dependent
interactions, responsible for the formation of the
magnetically-ordered state are much weaker than coulombic
interactions and are of order of T$_{c}$(T$_{N}$)/k$_{B}$, i.e.
of, say, 30 meV.

In QUASST the magnetic moment is associated with the open-shell
atoms. QUASST enables calculations of both the spin and orbital
moments \cite{20}. There is rapidly growing experimental evidence
for the existence of the orbital moment, thanks X-ray synchrotron
experiments, and QUASST enables its calculation using
well-established physical concepts. The used home-made computer
program BIREC for 3d is available on our web-side:
http://www.css-physics.edu.pl.

QUASST nicely illustrates that the whole magnetism is the
macroscopic quantum effect.

QUASST offers a solution of the Mott-insulator problem - an 3d
oxide is the insulator having bound electrons in the incomplete 3d
shell, that govern magnetic, electronic and spectroscopic
properties.

QUASST bridges the atomic physics and the solid-state physics.

In frame of QUASST low-energy excitations are neutral spin-like
excitations between CEF-like d states. According to QUASST these
neutral low-energy spin-like excitations, between two Kramers
conjugate states cause anomalous behavior of some Ce, Yb and U
compounds at low temperatures known as heavy-fermion phenomena.

Our approach is in agreement with a general idea of Mott that it
is strong electron correlations in 3d oxides that make electrons
in the incomplete 3d shell to stay rather localized than
itinerant. We are convinced that there is no other nowadays theory
of a solid containing transition-metal atoms that can describe
properties in so detailed and in so consistent way, both
zero-temperature properties and thermodynamics with the physically
clear distinction of the magnetically-ordered and paramagnetic
states.

The QUASST basic message can be defined also in a way that for the
physically adequate description of electronic and magnetic
properties of 3d-/4f-/5f-/atom compounds the CEF interactions,
completed with the intra-atomic-spin-orbit coupling, have to be
determined the first.

\subsection{The need for formulation of QUASST}

One can ask: ''Is this atomistic idea a new one in the solid-state
physics?'' Yes and no. No, as most of experimentalists naturally
discuss their results in terms of local properties. Yes, as
according to our knowledge none has been able to resist to
presently-in-fashion solid-state physics theories that simply
ignore the existence of the atom in a solid arguing that a solid
is so many-body object and that there are so strong intersite
correlations that the individuality of atoms is lost. In the
standard band-structure calculations the f and d electrons are
taken as itinerant forming a band. In the band there is a
continuum of the energy states within 1-5 eV. In our model there
are discrete states with energy separations even less than 1000
times smaller (1 meV, but 0 in case of Kramers ions). No, as there
are some text-books written about the crystal field, let mention a
book of Abragam and Bleaney \cite{27},of Ballhausen \cite{28} and
others \cite{29,30}. Yes, as they applied the CEF approach to some
diluted 3d systems, not to the concentrated ones, i.e. not to a
crystalline solid. Yes, as they have not been consequent enough
and they largely washed up the original idea. Yes, as there are
still very popular text-books and review articles that show very
schematic description of 3d states \cite{1,2,33,64} making use of
the single-electron picture. One should note that in the strong
CEF approach the n 3d electrons are treated as largely
independent, i.e. they do not form the strongly-correlated
3d$^{n}$ system in contrary to QUASST. QUASST corresponds to the
intermediate (3d) or weak (4f) crystal-field approach, but we
point out the fundamental importance of the intra-atomic
spin-orbit coupling, despite of its relative weakness for the 3d
ions. Also yes, as at present this atomic-like picture is
enormously prohibited in the leading physical journals and a
little is said about the discrete states at the magnetic and
annual strongly-correlated electron system (SCES) conferences.

We would like to add, preceding unfounded critics, that we do not
claim that everything can be explained only by single atoms but
our point is that the proper, i.e. physically adequate starting
point for the discussion of properties of a solid containing the
open-shell atoms is the consideration of its atomic-like states in
the local surroundings, i.e. in the lattice crystal field. Our
numerous computer experiments point out that e.g. the orbital
moment has to be unquenched in the solid-state physics of 3d-ion
containing compounds and our approach enables it. For instance, we
have derived the orbital moment in NiO to be 0.54 $\mu _{B}$ what
amounts to 20 \% of the total moment (2.53 $\mu _{B}$) \cite{38}.
Moreover, one should not consider our approach as the treatment of
an isolated atom - we start the discussion of NiO from the
consideration of the cation octahedra NiO$_{6} $ (more exactly -
the Ni$^{2+}$ ion in the octahedral crystal field). The whole NaCl
structure of NiO is built up from the edge sharing cation
octahedra. The perovskite structure, for instance, is built up
from the corner (and the edge) sharing cation octahedra along the
c direction (in the a-b plane). Thus, such octahedra cover the
whole macroscopic 3D body provided the perfect translational
symmetry. The CEF parameters contain information about the
interaction of the single ion with the whole charge surroundings.

Our approach is in agreement with the general conviction about the
importance of the electron correlations in description of
open-shell compounds - in our approach we start from the very
strongly-correlated limit in contrary to a weak correlation limit
of the LSDA approach.

For the better illustration of our point of view the reader is
asked to look into recent Phys. Rev. Lett. papers. In Ref.
\cite{65} authors, considering states of two 3d electrons of the
V$^{3+}$ ion in V$_{2}$O$_{3}$, came out with the continuous
electronic structure spread over 2.5 eV (Figs 2 and 3 of Ref.
\cite{65}). In Ref. \cite{66} two d electrons of the V$^{3+}$ ion
are considered to be largely independent. It shows that our quite
reasonable atomic-like approach is far from being accepted in
presently-in-fashion theoretical approaches. In Ref. \cite{67} the
continuous electronic structure for six 3d electrons in FeO
spreads over 8 eV (Fig. 8 of Ref. \cite{67}). In Refs
\cite{65,66,67} the orbital moment and the spin-orbit coupling is
completely ignored. These examples prove that our call for the
"unquenching" of the orbital moment and taking into account the
spin-orbit coupling is fully justified. We are convinced that in
all called papers d electrons form the crystal-field discrete
energy states with the importance of the s-o coupling. In FeO, in
the paramagnetic state, we expect a quite similar structure to
that presented in Fig. 8. Also magnetism develops analogically to
that discussed here for FeBr$_{2}$, Fig. 9, though FeO exhibits
much higher value of T$_{N}$. The electronic structure of the
V$^{3+}$ ion in V$_{2}$O$_{3}$ within the strongly-correlated
crystal-field approach has been derived by us in Ref. \cite{68}.

\section{Some further problems}

\subsection{CEF and QUASST \emph{vs} \emph{ab-initio} calculations}

Numerous successes of the CEF approach indicate that so subtle
effect as charge quadrupolar and octupolar interactions, with
energies smaller than 1 meV, have to be incorporated into each
modern solid-state theory. In QUASST the real space description is
used in contrast to the reciprocal space used in band theories.

Many-electron states and many-electron wavefunctions discussed in
CEF and in QUASST contrast single-electron states and
single-electron wavefunctions considered in the band theories.
They are also completely different from single-electron CEF
d-orbitals $t_{2g}$ and $e_{g}$. In general, the wavefunction and
the energy of the whole atomic-like 3d$^{n}$ many-electron system
can be constructed from single-electron wavefunctions of
d-orbitals [$l$,$l_{z}$], with $l$ = 2, and $l_{z}$ = $\pm{2}$,
$\pm{1}$, and 0. Undoubt successes of the CEF and QUASST
description of many 3d-/4f-/5f-atom compounds indicate that this
construction from single-electron wavefunctions {$l$,$l_{z}$} have
to be oriented in order to reproduce the atomic-like eigenstates
and wavefunctions of the desired ion. Other words, ambitious
solid-state theories starting the description of a 3d solid from a
(gaseous) jellium of independent electrons must put such
conditions and such potentials and mutual electron interactions
into their theories in order to get 3d atoms or 3d ions and to
reproduce the spectral term structure, known from the atomic
physics. In QUASST we take the existence of atoms or ions as the
physical fact that is not necessary to prove. Their existence in
the reality confirms the validity of our assumption. We treat the
problem of the description of atoms or ions, in fact the
description of the term electronic structure, as the subject of
the atomic physics.

The preservation in a solid of the atomic-like 3d$^{n}$
configuration is the manifestation of strong electron
correlations. The developed by us the QUASST theory is, according
to us, the physical realization of the general idea of Sir Nevill
Mott that it is strong electron correlations that make electrons
in the incomplete 3d shell to stay rather localized than itinerant
in 3d oxides. The 3d oxides are known at present as Mott
insulators.

Crystal-field parameters can be, in general, calculated \emph{ab
initio} as the multipolar potentials of the total charge
distribution in a solid. The total charge distribution is related
to all charges in a solid; all nuclei and all electrons. This
first-glance hopeless task becomes, however, tractable in a
crystalline solid thanks the translational and local symmetry.
Such calculations, direct \emph{ab initio} electrostatic
multipolar potentials of the 3D surroundings, have been
demonstrated by Hutchings \cite{69} by considering point charges
of the octahedral and tetrahedral symmetries. Calculations by
Hutchings concern the crystal field for rare-earth ions but the
calculations are the same for 3d ions if one replaces J by L
quantum number. For the octahedral symmetry of the 3d-ion cation
in the oxygen octahedron the CEF coefficient A$_{4}$ can be
calculated \emph{ab initio} by consideration of electrostatics
like Eqn 6. The single-ion prediction of the variation of the
B$_{4}$ parameter enables study of CEF interactions across the
whole isostructural series and somehow to predict physical
properties of a compound knowing the local symmetry and the
involved cation. The reversal of the electronic structure for the
Ni$^{2+}$ and Co$^{2+}$ ions ($d^{7}$ and $d^{8}$ configurations)
have been noticed already by Van Vleck \cite{59}.

Although the calculated \emph{ab initio} value of the B$_{4}$
parameter in LaCoO$_{3}$ for the nearest oxygen octahedron, see
section 4.2, is eight times smaller than the recent evaluation of
260 K it is extremely important that this \emph{ab initio}
calculation provides the proper sign of B$_{4}$ parameter because
this sign determines the ground state in the oxygen octahedron.
There is a hope that taking into calculations all charges and
other physical effects bridges both values. However, we point out
that by the recent experimental evaluation of the strength of
crystal-field interactions it turns out that the commonly used
strong crystal-field approach, with breaking the atomic-like
correlations, is not justified. In fact, turns out to be useless.

In fact, QUASST is indeed \emph{ab initio} theory - it takes
profit of the atomic physics with the spectral term electronic
structure and later \emph{ab initio} calculates the electronic
structure of the given magnetic ion in a crystalline solid, taking
into account the crystal field potentials and the spin-orbit
coupling. The electronic structure from \emph{first principles} by
means of the statistical physics provides both zero-temperatures
properties and thermodynamics.

We think that the largest error in the \emph{ab initio} CEF
calculations of $B_{4}$, mentioned in section 4.2, is associated
with the determination of the charge distribution in the crystal
and the theoretical evaluation of the ionic octupolar moment of
the 3d cation, in particular with the value of $<r^4>$. Our
preliminary studies indicate that there is no strong dependence of
$<r^4>$ across the 3d series as values in Fig. 7 show.

\subsection{One-term \emph{vs} all-terms fine electronic structure}

Of course, the QUASST calculations should be performed within
all-terms space involving the ${10 \choose n}$ states allowed by
the Pauli principle. Of course, it would be better to take into
consideration all interactions known to exist in a solid. However,
more parameters more difficult to get physically adequate results
is. Physics is the science of approximations - the most important
is to catch the most essential physical ingredient(s) and made
only physically proper approximations. We are convinced that
\textbf{by taking into account the atomic physics and the
spin-orbit coupling we have caught the most essential physical
ingredients. Surely we have the physically adequate number of
states involved in the low-energy electronic structure} like it
was shown for FeBr$_{2}$. For the V$^{4+}$ ion in
Na$_{2}$V$_{3}$O$_{7}$ there is only one term, $^{2}$D. Thus
calculations for the d$^{1}$ configuration for the V$^{4+}$ and
Ti$^{3+}$ ions are equivalent to all-terms calculations. The
performing all-terms calculations, provided that there is no term
reorganization as occurs in LaCoO$_{3}$, will lead to only small
changes in energies and eigenfunctions compared to Hund's rules
ground term calculations, but the number of electronic states will
be preserved. It is important as the number of available
low-energy states determines the overall behavior of
thermodynamical properties, like the total entropy. In LaCoO$_{3}$
the subterm reorganization occurs and the $^{1}$A$_{1}$ subterm
becomes the ground subterm, but it proceeds in the
fully-controlled and in the fully-understood way.

Thus, all-terms calculations do not introduce changes in the
physical description of properties in comparison to the one-term
calculation. It is in contrast to the revolutionary change of
physics going from the orbital-only states to the spin-orbit
states discussed in the strongly-correlated crystal-field approach
and in QUASST. But it is Nature that chooses the physically
adequate approach.

\subsection{Towards 4d and 5d atom containing compounds}

We hope that the discrete states also are present in 4d-/5d-atom
containing compounds. Calculated electronic structure for the
Ru$^{4+}$ ion in Sr$_{2}$RuO$_{4}$ has been presented in Ref.
\cite{70}. In 4d and 5d ions the spin-orbit coupling is much
stronger that in case of 3d ions. The j-j coupling should be
rather more applicable than the Russell-Saunders coupling used for
the description of 3d ions. The substantial applicability of the
strongly-correlated CEF approach to 5f-atom compounds have been
demonstrated for UPd$_{2}$Al$_{3}$, UGa$_{2}$ and NpGa$_{2}$
\cite{25,71}. Magnetic and electronic properties of
UPd$_{2}$Al$_{3}$ and UGa$_{2}$ turn out to be related to the
5$f^{3}$ configuration like in the U$^{3+}$ ion.

\subsection{Testing of the quantum mechanics}

Accepting the CEF and QUASST theory one can say that the
crystalline solid provides the wide possibility to study the Stark
and Zeeman effect on paramagnetic ions, their discrete states, in
very complicated multipolar electrostatic fields and in extremal
magnetic fields, unavailable presently in the laboratory.

In deriving the states, their energy and eigenfunction, an
important problem of the atomic physics, the applicability of the
Russell-Saunders or j-j coupling can be effectively studied. From
this point of view the perfect description of the g tensor of the
Co$^{3+}$ ion \cite{18} and the behavior of its localized states
observed in the ESR experiment on single crystalline LaCoO$_{3}$
in magnetic fields up to 30 T along different crystallographic
directions we take to be of a remarkable scientific importance.

It is also very important from the unification point of view that
the single-ion Hamiltonian widely used in the analysis of
electron-paramagnetic resonance spectra of 3d-ion doped systems
has been successfully used for systems, where the 3d atoms are the
full part of the crystal.

\section{CONCLUSIONS}

We have presented the crystal-field approach with strong electron
correlations, extended to the Quantum Atomistic Solid-State theory
(QUASST), as a physically relevant theoretical model for the
description of electronic and magnetic properties of 3d-atom
compounds. Its applicability has been illustrated for LaCoO$_{3}$,
FeBr$_{2}$ and Na$_{2}$V$_{3}$O$_{7}$. According to the QUASST
theory in compounds containing open 3$d$-/4$f$-/5$f$-shell atoms
the discrete atomic-like low-energy electronic structure survives
also when the 3d atom becomes the full part of a solid matter.
This low-energy atomic-like electronic structure, being determined
by local crystal-field interactions and the intra-atomic
spin-orbit coupling, predominantly determines electronic and
magnetic properties of the whole compound. It should not mean that
other mechanisms than CEF and the spin-orbit coupling are not
present in a solid, but we are convinced that analysis of
electronic and magnetic properties of 3d-/4f-/5f-atom compounds is
necessary to start from the determination of the crystal-field
interactions and of the localized CEF-like states.

We understand our theoretical research as a continuation of the
Van Vleck's studies on the localized magnetism. We point out,
however, the importance of the orbital magnetism and the
intra-atomic spin-orbit coupling for the physically adequate
description of real 3d-ion compounds. Our studies clearly indicate
that it is the highest time to ''unquench'' the orbital moment in
solid-state physics in description of 3d-atom containing
compounds.

\end{document}